\documentclass[a4paper,11pt]{article}
\usepackage{jinstpub} 
\usepackage{lineno}
\usepackage{wasysym}
\usepackage{bm}
\usepackage{threeparttable}
\usepackage[export]{adjustbox}

\title{\boldmath Characterization of silicon photomultipliers for their application in muon scattering tomography}

\author[a,b]{Binghao Sun,}
\author[b,*]{Huiling Li\note[*]{Corresponding author},}	
\author[c]{Quanyin Li,}
\author[b]{Hui Liang,}
\author[b]{Cong Liu,}
\author[b]{Hongbo Wang,}
\author[c,b]{Zibing Wu,}
\author[b]{Suyu Xiao,}
\author[c,b]{Weiwei Xu}

\affiliation[a]{University of Jinan, 336 Nanxinzhuang West Road, Jinan, China}
\affiliation[b]{Shandong Institute of Advanced Technology, 1501 Panlong Road, Jinan, China}
\affiliation[c]{Shandong University, 27 Shanda Nanlu, Jinan, China}
\emailAdd{huiling.li@iat.cn}
\abstract{Muon scattering tomography is a non-destructive technique used to image different materials by utilizing natural cosmic ray muons. Typically it requires position-sensitive detectors with a sub-millimeter resolution to effectively distinguish high-$Z$ materials in a compact system. The plastic scintillating fiber detector is a feasible candidate and is currently being designed with one-dimensional silicon photomultiplier (SiPM) readout. In this work, we constructed experimental setups to characterize three different SiPMs from the NDL, SensL, and HPK manufacturers for optimal performance of the scintillating fiber detector. The breakdown voltage, temperature compensation factor, dark noise, and photodetection efficiency of each SiPM are evaluated and summarized. Among the SiPMs tested, the HPK SiPM demonstrated the lowest dark count rate and crosstalk probability while exhibiting the best photodetection efficiency response at the emission wavelengths of the scintillating fibers. This makes the HPK SiPM particularly well-suited to meet the requirements of the detector and serves as a reference for further customization of the one-dimensional SiPM array.}

\keywords{muon scattering tomography, silicon photomultiplier, breakdown voltage, temperature compensation, dark noise, photon detection efficiency}

\begin{document}
\maketitle
\flushbottom

\section{Introduction}
The muon scattering tomography (MST) technique can image dense objects non-destructively by utilizing highly penetrating cosmic ray muons. This method is based on the principle of multiple Coulomb scattering of cosmic ray muons crossing different materials. A typical MST facility comprises two tracking detectors placed before and after the target to register these deviation angles. Since its introduction by Los Alamos National Laboratory in 2003~\cite{Borozdin2003}, various research groups have employed different types of position-sensitive detectors in MST systems, including nuclear emulsions, plastic scintillators, and micro-pattern gaseous detectors~\cite{Bonomi:2020}. Two key parameters in optimizing the imaging time and accuracy of these detectors are their spatial resolution and active area. 

\begin{figure}[!h]
\centering
\hspace{-1cm}
\includegraphics[scale=0.45]{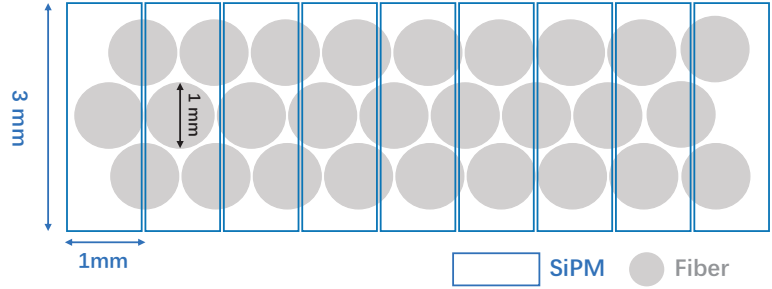}
\vspace{-1.8cm}
\caption{The compact SciFi detector module is designed with staggered plastic scintillating fibers and one-dimensional SiPM readout.}
\label{fig:scifi}
\end{figure}

The plastic scintillating fiber (SciFi) detector~\cite{Gruber:2020shw} with one-dimensional highly segmented silicon photomultiplier (SiPM) readout can achieve much finer spatial resolution compared to conventional bulk plastic scintillator detectors~\cite{Anghel:2012,Riggi:2018}. Thanks to the long attenuation length and high light yield of scintillating fibers, SciFi can be developed with a large area while still maintaining a high spatial resolution. This characteristic makes it a suitable candidate to construct a more compact MST system. To develop a SciFi detector capable of a submillimeter spatial resolution and detection efficiency exceeding 95\%, we designed the structure of a SciFi module (see Fig.~\ref{fig:scifi}) with staggered fibers of 1 mm in diameter and one-dimensional SiPM array with a channel size of $1\times3$ mm, as proposed in our design of an MST system to identify both low-$Z$ and high-$Z$ materials~\cite{Chen:2023}.

SiPM is a solid-state photon sensor made up of multiple small avalanche photodiodes that operate in Geiger mode~\cite{Acerbi:2019, Piemonte:2019, Klanner:2019}. Compared to conventional photomultiplier tubes, SiPMs are compact, insensitive to magnetic fields, provide high internal gain, operate at low voltages, and exhibit good photon detection efficiency (PDE) as well as time resolution. Its major disadvantages are the high dark noise rate and temperature dependence, which may render fluctuations of SiPM-based detector capabilities, especially at low signal levels. A typical cosmic ray muon with an energy of 4 GeV passing through a SciFi module, which has a light yield of 8000 photons per MeV and a trapping efficiency of 5.4\%, will produce up to approximately 200 photons on average that reach the surfaces of the SiPMs. To effectively detect photons from scintillating fibers, the SiPM is required to have a high PDE sensitive to light in the wavelength range of 400 nm to 700 nm, appropriate cell pitches to ensure a linear response range for at least 200 photons, and a dark noise rate below $\rm{3\times10^{6}}$ $\rm{kHz/mm^{2}}$.

Currently, there are no SiPM products available that match the required channel size for the SciFi module. To optimize the performance of the SciFi detector and determine the appropriate parameters for further SiPM customization, we characterize and compare three advanced SiPMs from HPK, SensL, and NDL manufacturers in this work. The following sections evaluate the breakdown voltage and temperature compensation factors for each SiPM to maintain relatively stable gain at different temperatures. Additionally, dark noise levels and PDE are crucial for ensuring good spatial resolution and high detection efficiency of the SciFi detector, and these parameters are also assessed using test setups in our laboratory.

\section{SiPM devices}
\begin{table}[htbp]
\begin{threeparttable}
\centering
\caption{Some parameters of three SiPMs provided by their vendors.}
\smallskip
\begin{tabular}{cccccc}
\hline 
Parameters \tnote{1} & S13363-3050NE-16 & MicroFJ-30035-TSV & EQR15 11-3030D-S \\
\hline
Manufacture & HPK & SensL & NDL \\
Active area/channel [mm] & 3$\times$3 & 3.07$\times$3.07 & 3$\times$3 \\
Cell pitch [$\mu$m] & 50 & 35 & 15 \\
Number of cells & 3585 & 5676 &  39996 \\
Fill factor [\%] & 74 & 75 & - \tnote{2}\\
Breakdown voltage $\rm{V_{bd}}$ [V] & 53$\pm$5 & 24.2-24.7& 30 \\
Operating voltage & $\rm{V_{bd}+3}$ & $\rm{V_{bd}+1}-\rm{V_{bd}+6}$ &$\rm{V_{bd}+8}$\\ 
Terminal capacitance [pF] & 320 & 1070 \tnote{3} & 50 \\
\hline
\end{tabular}
\label{tab:sipm}
\begin{tablenotes}
\item[1] Measurements made at 25${\rm^{o}C}$ for HPK, 21$\rm{^{o}C}$ for SensL and 20$\rm{^{o}C}$ for NDL SiPM.
\item[2] Not available in NDL SiPM datasheet.
\item[3] Anode capacitance for SensL SiPM.
\end{tablenotes}
\end{threeparttable}
\end{table}	
A SiPM is a semiconductor-based device consisting of multiple microcells connected in parallel, where each cell is a series connection of an avalanche photodiode and a quenching resistor. It can respond to a single photon, generating a current pulse with a width of several tens of nanoseconds. In this work, we selected three advanced SiPMs sensitive to emitted photons in the 400 nm to 700 nm range from a scintillating fiber~\cite{Cavalcante:2018}. They are S13363-3050NE-16 from HPK~\cite{HPK:SiPM}, MicroFJ-30035-TSV from SensL~\cite{SensL:SiPM} and EQR15 11-3030D-S from NDL~\cite{NDL:2019}, and part of their parameters from vendors as listed in Tab.~\ref{tab:sipm}.
\begin{itemize}
\item The HPK SiPM is a one-dimensional array of 16 channels fabricated using advanced Through Silicon Via (TSV)~\cite{Parellada:2023} and chip size package technologies. The TSV technology significantly reduces the gap around the outer periphery of one channel to 0.2 mm on all four sides, compared to the 0.4 mm gap found in HPK wire-bonded products such as the S13360-3050PE. One of the 16 channels is used for characteraction in the following.
\item The SensL SiPM also uses a high-performance TSV packaging process to reduce dead area. It includes a standard output from the anode and a capacitively coupled fast output. Only the standard output is tested in this work. 
\item The NDL SiPM employs an intrinsic epitaxial layer as the quenching resistors (EQR) and utilizes a continuous silicon cap layer as an anode to connect all the cells, which allows larger microcell density while retaining a high PDE. 
\end{itemize}
These three SiPMs have different cell pitches as listed in Tab.~\ref{tab:sipm}. Their cell pitches can provide a linear response range to at least 200 photons with non-linearity less than 5\% in one SiPM channel of Fig.~\ref{fig:scifi}. This is evaluated with the formula $\rm{N_{fired}=N_{cell}(1-e^{-N_{ph}\frac{PDE}{N_{cell}}})}$~\cite{Acerbi:2019}, assuming a large PDE value of 0.6. 

\section{Test setup}\label{sec:setup}
\begin{figure}[!t]
\begin{center}
\includegraphics[scale=0.5]{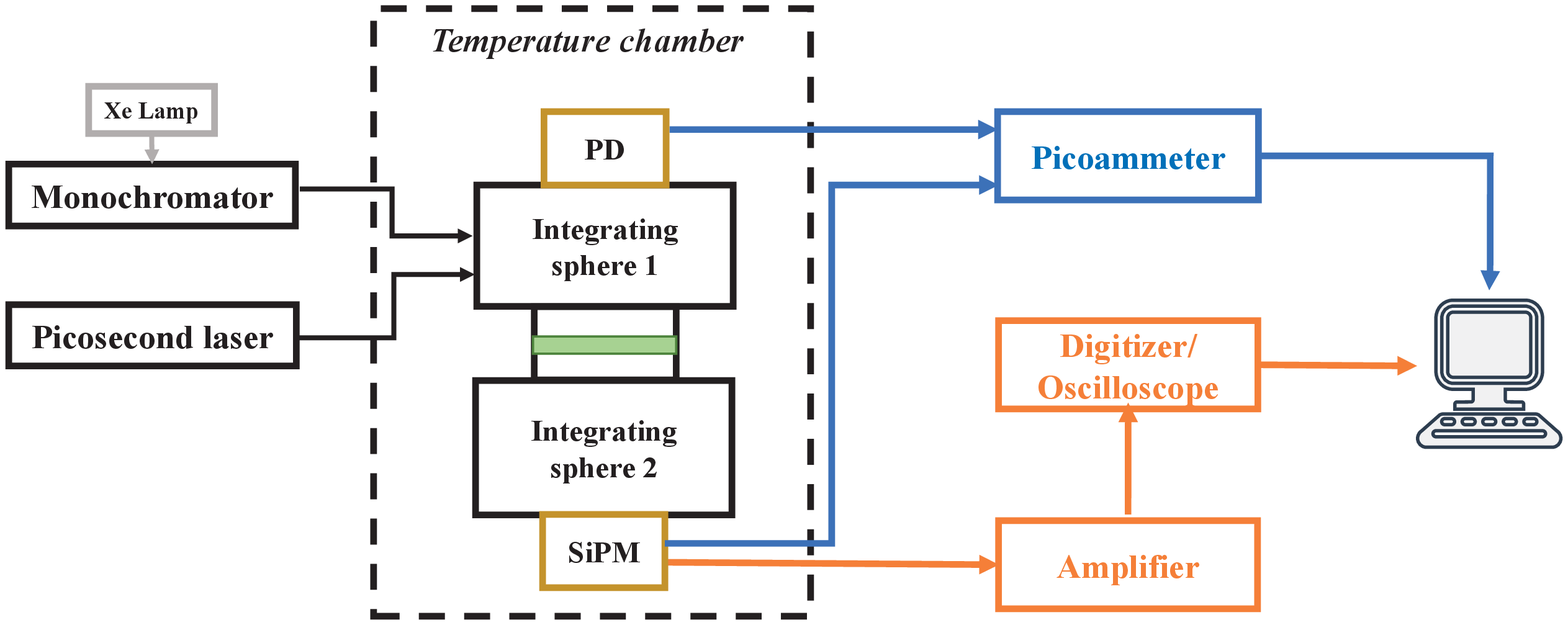} 
\end{center}
\vspace{-3cm}
\caption{Schematic diagram of test setup for SiPM characterization.}
\label{fig:setup}
\end{figure}

\begin{figure}[!t]
\begin{center}
\begin{tabular}{l}
\includegraphics[width=0.2\textwidth]{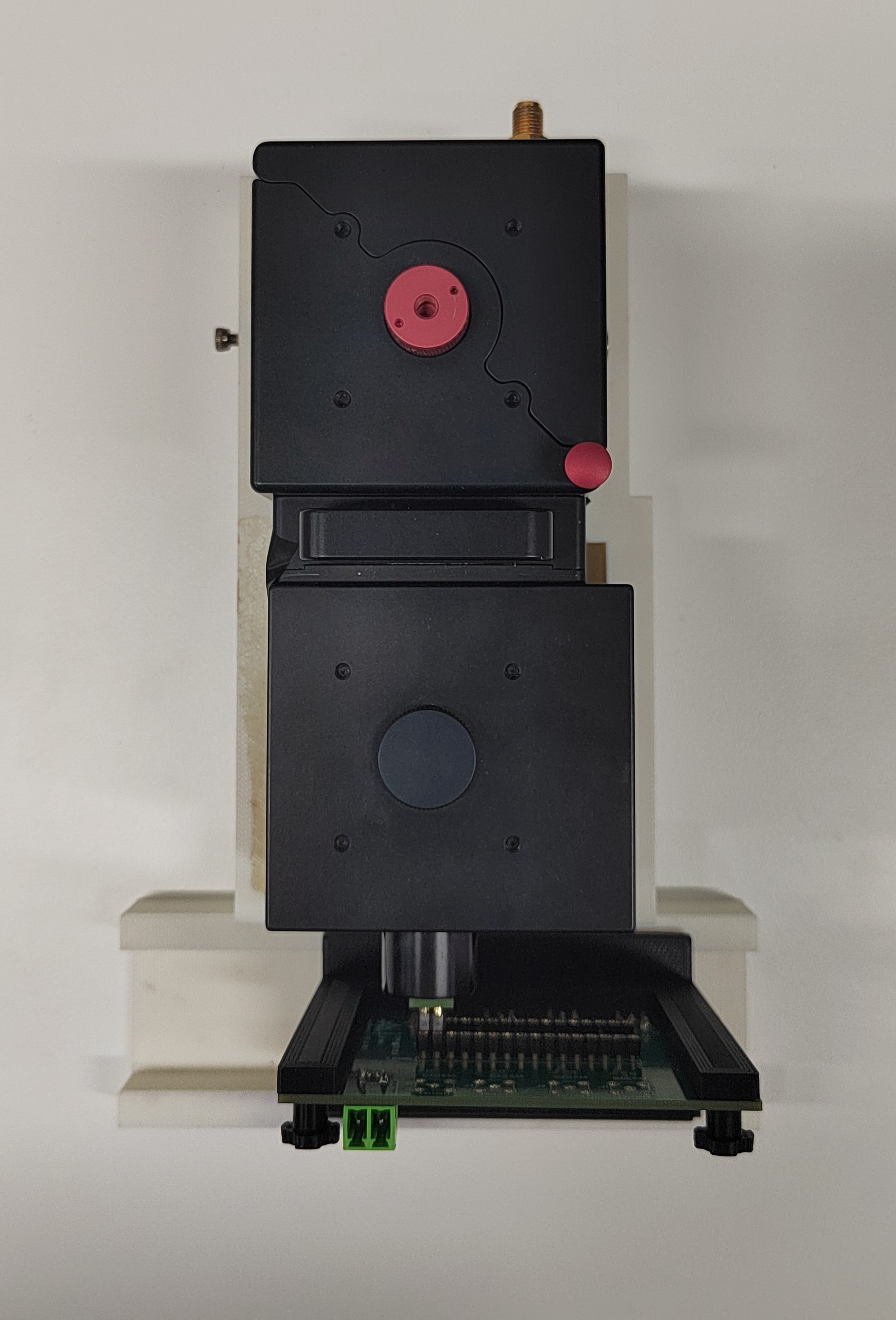}
\hspace{0.7cm}
\includegraphics[width=0.2\textwidth]{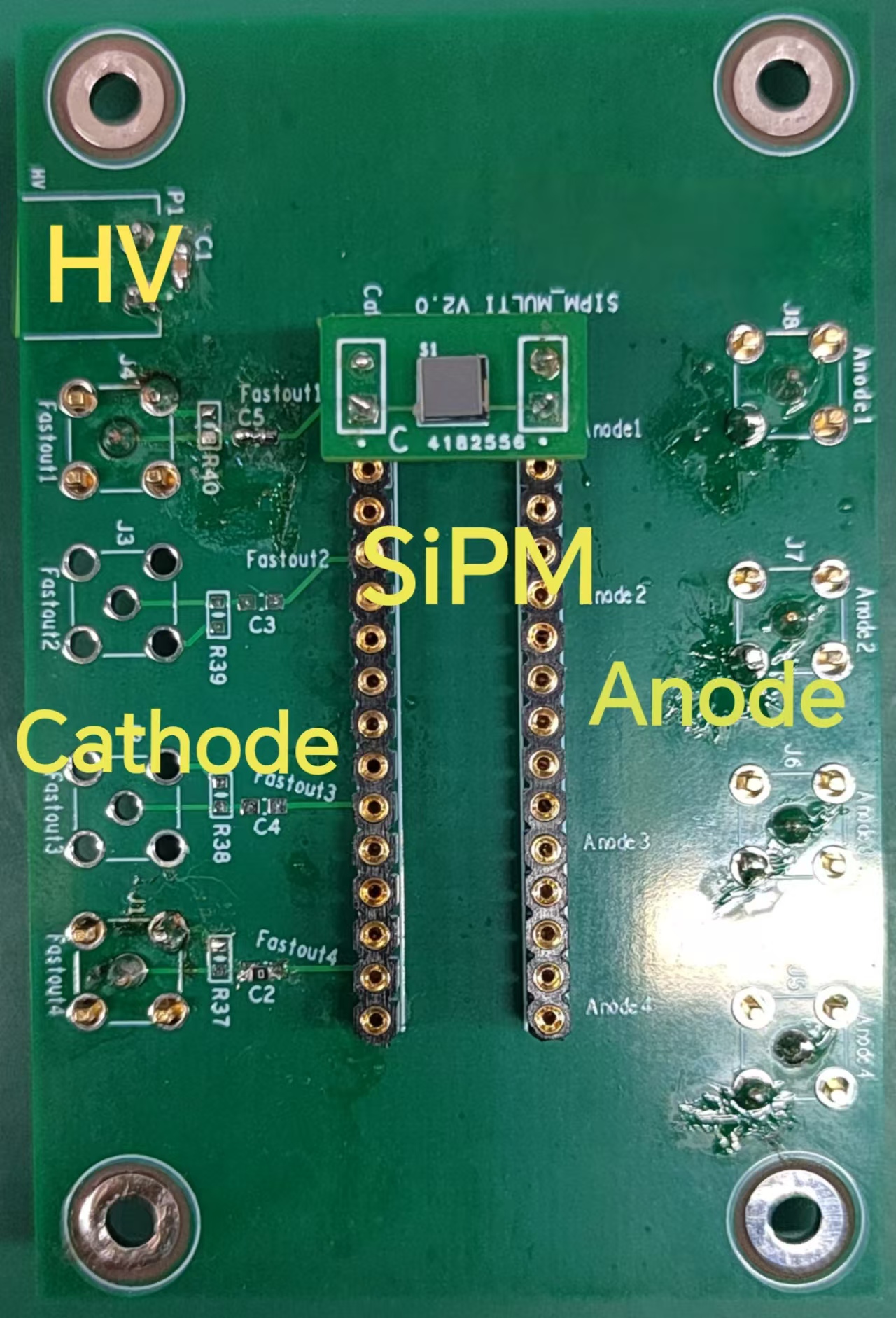}
\hspace{0.3cm} 
\includegraphics[width=0.42\textwidth]{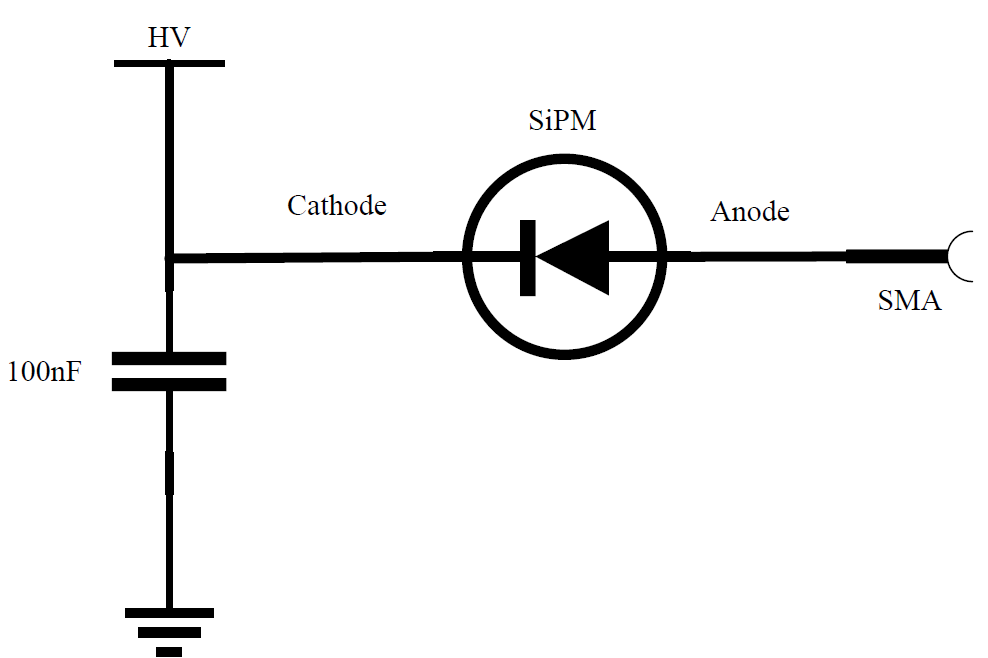} 
\end{tabular}
\end{center}
\vspace{-0.6cm}
\caption{(Left) The light-tight connected device of two integrating spheres with PD and SiPM attached. (Middle) The common electrical connection board and (Right) the corresponding schematic of three SiPMs, where HV is connected with the power supply and SiPM signals from the anode are sent out to either amplifier or picoammeter. }
\label{fig:SphereCircuit}
\end{figure}

A test setup for SiPM characterization is built with its schematic diagram shown in Fig.~\ref{fig:setup}. 
A temperature chamber with a range of $\rm{-40^{o}C}$ to $\rm{150^{o}{\rm C}}$ is used to maintain a SiPM operating at various temperatures and also serves as a dark box covered with black shielding to prevent ambient light from entering. To accommodate the dynamic range of the SiPM, a light-tight connected device consisting of two $\diameter 50 {\rm mm}$ integrating spheres from Thorlabs, as shown in the left of Fig.~\ref{fig:SphereCircuit}, is implemented to diffuse incident light and adjust photon levels onto the SiPM surfaces. A common electrical connection board for three SiPMs is designed, as illustrated in Fig.~\ref{fig:SphereCircuit}, and is utilized for subsequent measurements. The SiPM operates under reverse bias supplied by a Keithley DC power supply, and its signal from the anode is sent out to an amplifier or a picoammeter. All data acquisition equipment is located outside the temperature chamber. This setup enables the characterization of the breakdown voltage, gain, dark noise, and photon detection efficiency (PDE) of the SiPM.
\begin{itemize}
\item \textbf{For breakdown voltage}: The SiPM is directly connected to the Keithley DC power supply and a picoammeter from the Keithley 6482 series, of which the maximum resolution at 2 nA range is 1fA and the accuracy is $\rm{1\%+2 pA}$ at $\rm{23^{o}C\pm5^{o}C}$. The current-voltage (I-V) characteristics are automatically collected on a computer in dark conditions at various temperatures. 
\item \textbf{For gain measurement}: The pulsed light directed at the SiPM is provided by an NKT picosecond pulsed diode laser PIL040-FC with a fixed wavelength of 405 nm and diffused in the integrating sphere. SiPM signals are first amplified by a factor of 10 using the CAEN N979 module and then triggered into a digitizer DT5742 with 12-bit ADC for charge integral. To estimate the absolute gain value, a Tektronix MSO5 oscilloscope with a 6.25 GS/s sampling rate and 13-bit resolution is then used to register waveforms for charge calibration. 
\item \textbf{For dark noise}: The SiPM operates in a dark environment , and its waveforms after amplification by N979 are recorded by the oscilloscope with 1.25GS/s sampling rate for offline analysis.
\item \textbf{For PDE measurement}: A continuous light source provided by a 500 W Xenon lamp and a monochromator with a bandwidth from 200 nm to 2500 nm is guided into the integrating sphere by a fiber. The light intensity is monitored by a calibrated photodiode (PD) attached to the integrating sphere. SiPM under test is positioned at another port of the integrating sphere to detect the diffused photons. Currents from both the PD and the SiPM are aquisited synchronously by the picoammeter.
\end{itemize}

\section{Characterization methods and results}\label{sec:method}
\subsection{Breakdown voltage and temperature compensation}
Each microcell of a SiPM operates in Geiger mode by applying a reverse bias voltage ($V_{\rm {bias}}$) larger than its breakdown voltage ($V_{\rm {bd}}$). When biased just above the breakdown voltage, a sufficiently high electric field is induced within a cell that enables a single charge carrier to be injected into the depletion layer to trigger a self-sustaining avalanche. Therefore, the breakdown voltage can be identified where the I-V curve of the SiPM starts to rise very deeply. Methods of deriving the breakdown voltage from I-V curves of a SiPM are summarized in Ref.~\cite{Nagy:2016mfv}. In this context, the relative derivative method is employed, in which the breakdown voltage corresponds to the peak of the logarithmic derivative of the current. 

\begin{figure}[!t]
\begin{center}
\begin{tabular}{l}
\hspace{-0.3cm}
\includegraphics[width=0.55\textwidth]{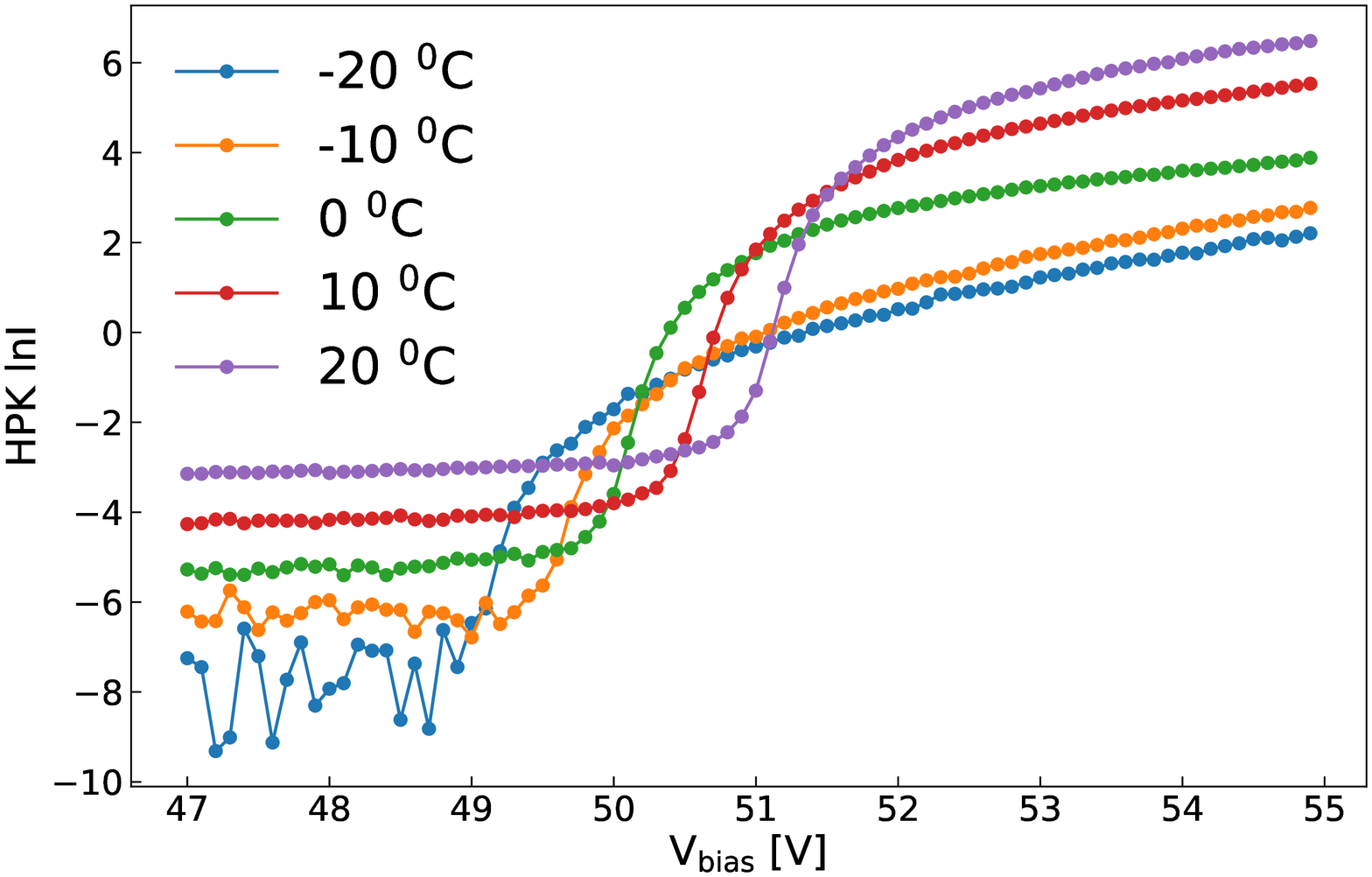}
\hspace{-1cm}
\includegraphics[width=0.55\textwidth]{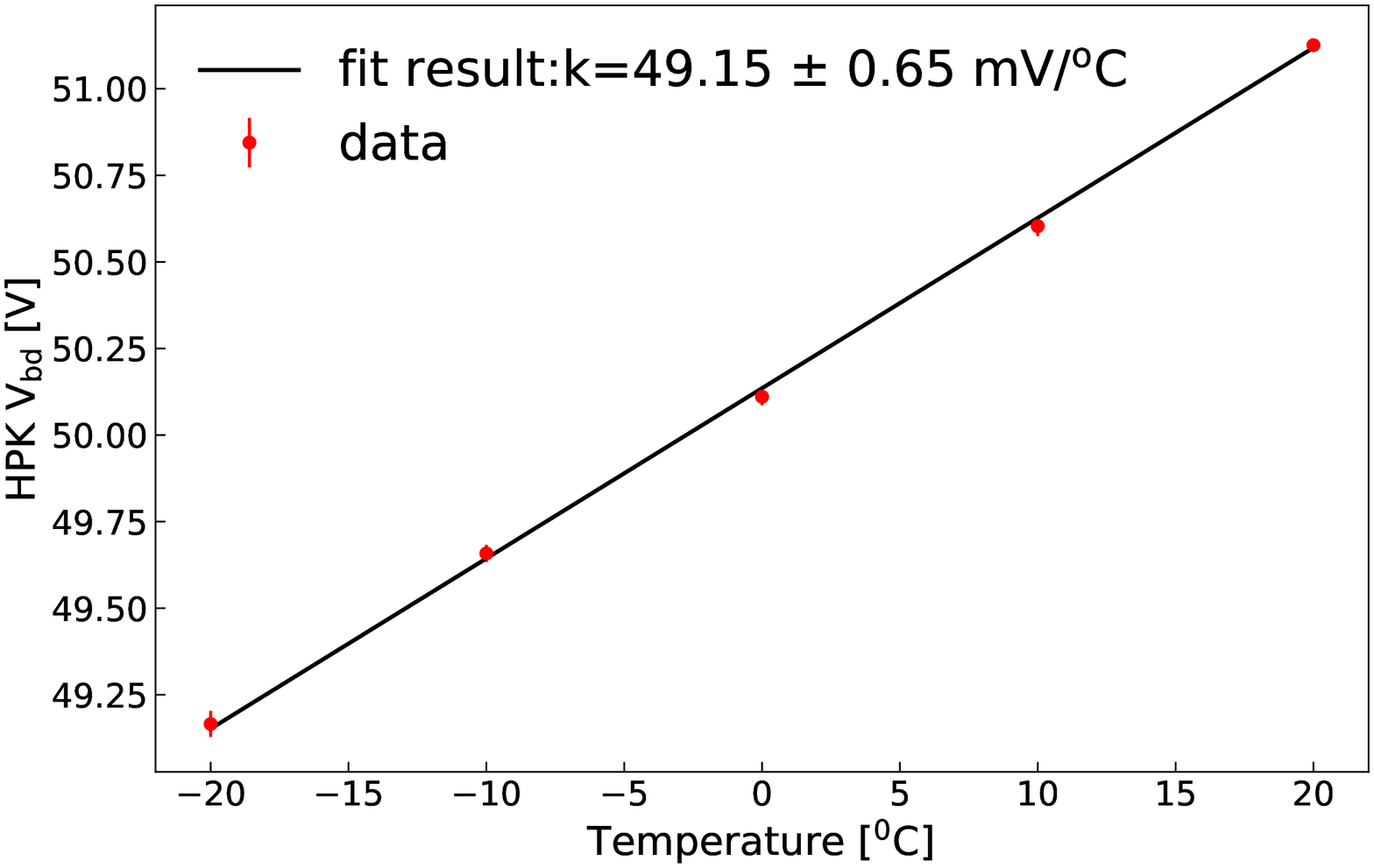} \\
\hspace{-0.3cm}
\includegraphics[width=0.55\textwidth]{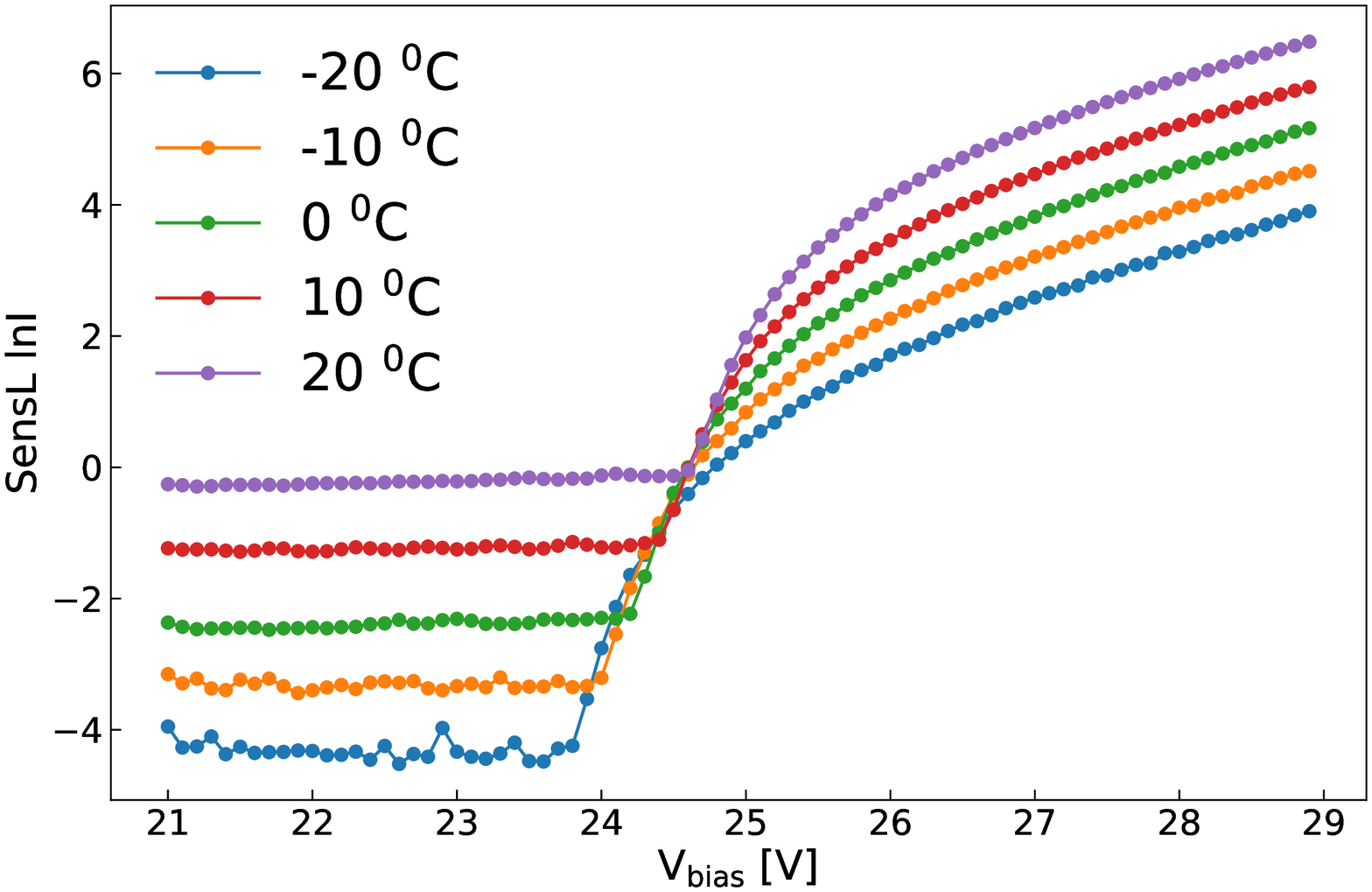}
\hspace{-1cm}
\includegraphics[width=0.55\textwidth]{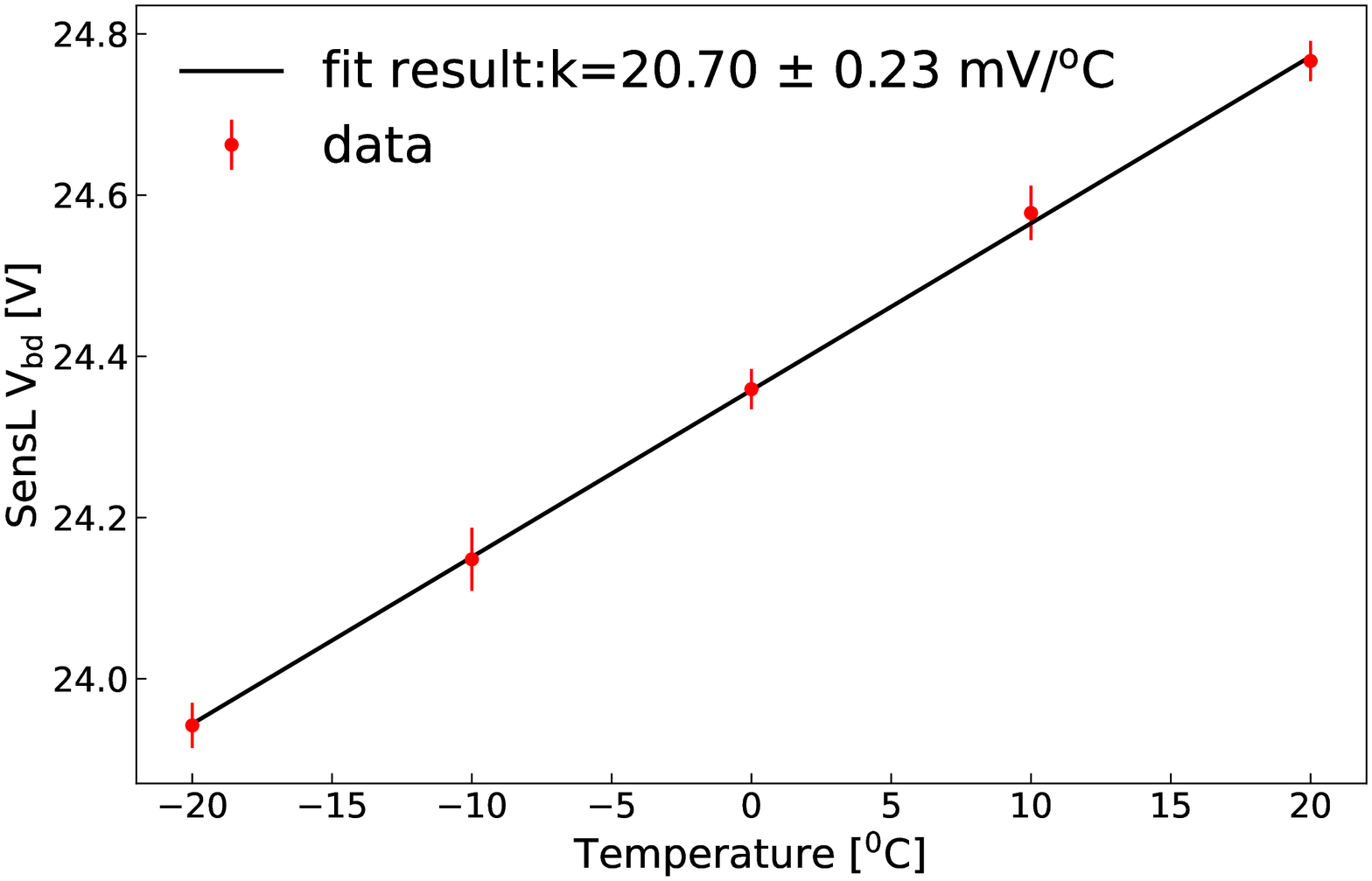} \\
\hspace{-0.3cm}
\includegraphics[width=0.55\textwidth]{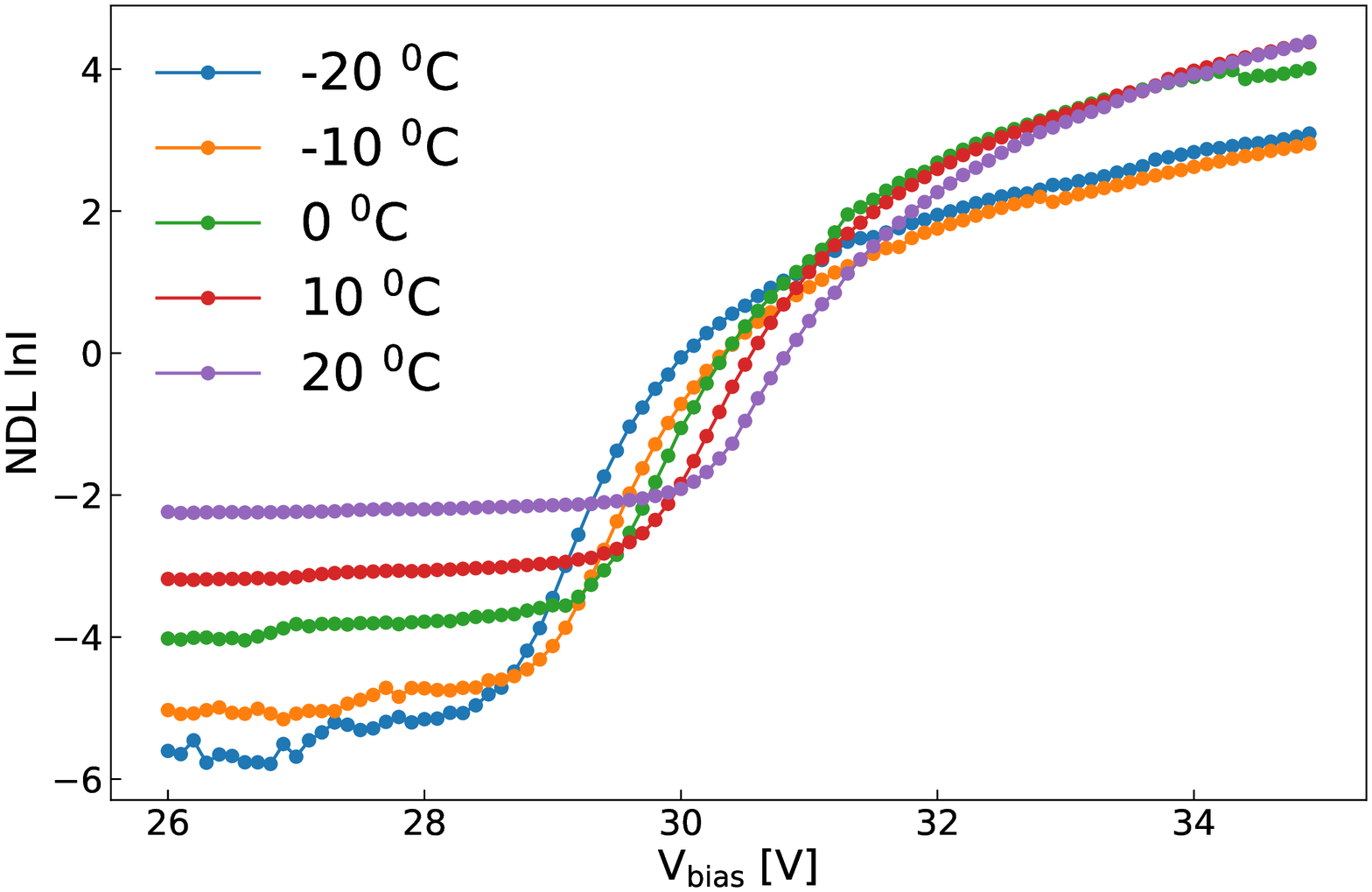}
\hspace{-1cm}
\includegraphics[width=0.55\textwidth]{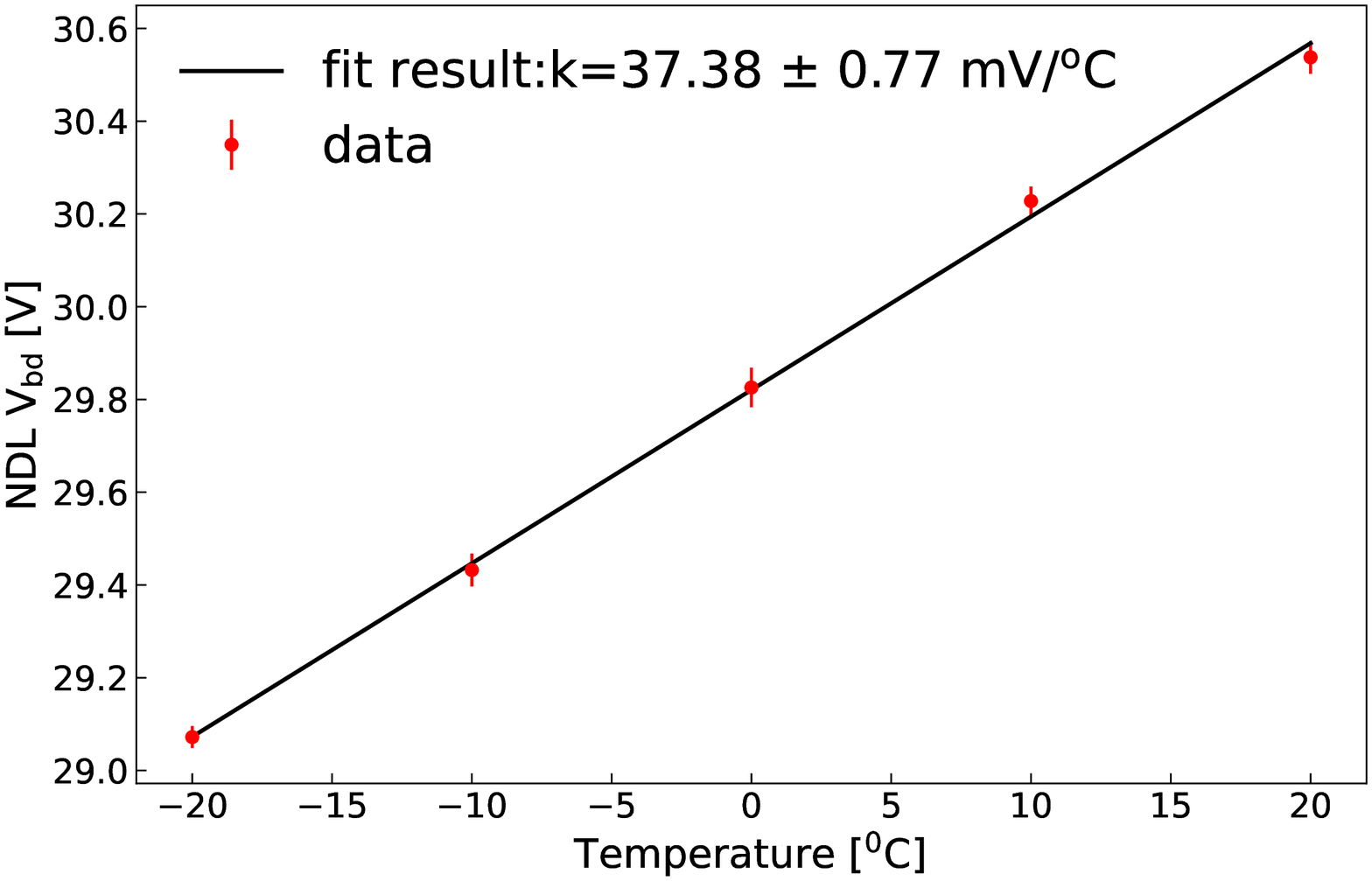} \\
\end{tabular}
\end{center}
\vspace{-0.8cm}
\caption{(Left) I-V curves  and (Right) the corresponding breakdown voltages at different temperatures of HPK, SensL and NDL SiPM in dark conditions.}
\label{fig:Vbd}
\end{figure}

In the left column of  Fig.~\ref{fig:Vbd}, I-V curves of HPK, SensL, and NDL SiPM are acquired at 20, 10, 0, -10, -20$\rm{^{o}C}$ under dark conditions. The fluctuations in dark current at -20$\rm{^{o}C}$ result from the resolution limit of the picoammeter. The corresponding breakdown voltages derived from the maximum of $dlnI/dV$ are indicated with red points in the right column of Fig.~\ref{fig:Vbd}. It shows that $V_{\rm {bd}}$ increases with the ambient temperature and fitted with a straight line:
\begin{equation}
V_{\rm {bd}}(T)=kT+b\;
\label{eq:bdT}
\end{equation}
where $k$ is called the temperature compensation factor and is related to the width of the depletion region in a microcell. If $V_{\rm {bias}}$ of a SiPM is adjusted with ambient temperature using this compensation factor, the SiPM gain $G=C\left(V_{\rm{bias}}-V_{\rm{bd}}\right)$ can be maintained relatively stable.

\begin{figure}[!h]
\centering
\includegraphics[scale=0.23]{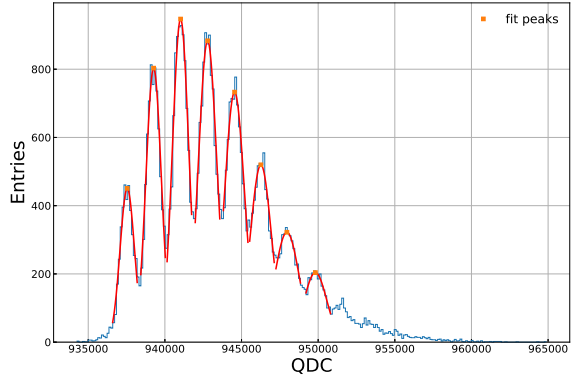}
\vspace{-0.5cm}
\caption{The charge spectrum of HPK SiPM operated at 54V biased voltage and 20 $\rm{^{o}C}$, where each peak is fitted with a Gaussian distribution. }
\label{fig:QDC}
\end{figure}

\begin{figure}[!t]
\begin{center}
\begin{tabular}{l}
\hspace{-0.5cm}
\includegraphics[width=0.55\textwidth]{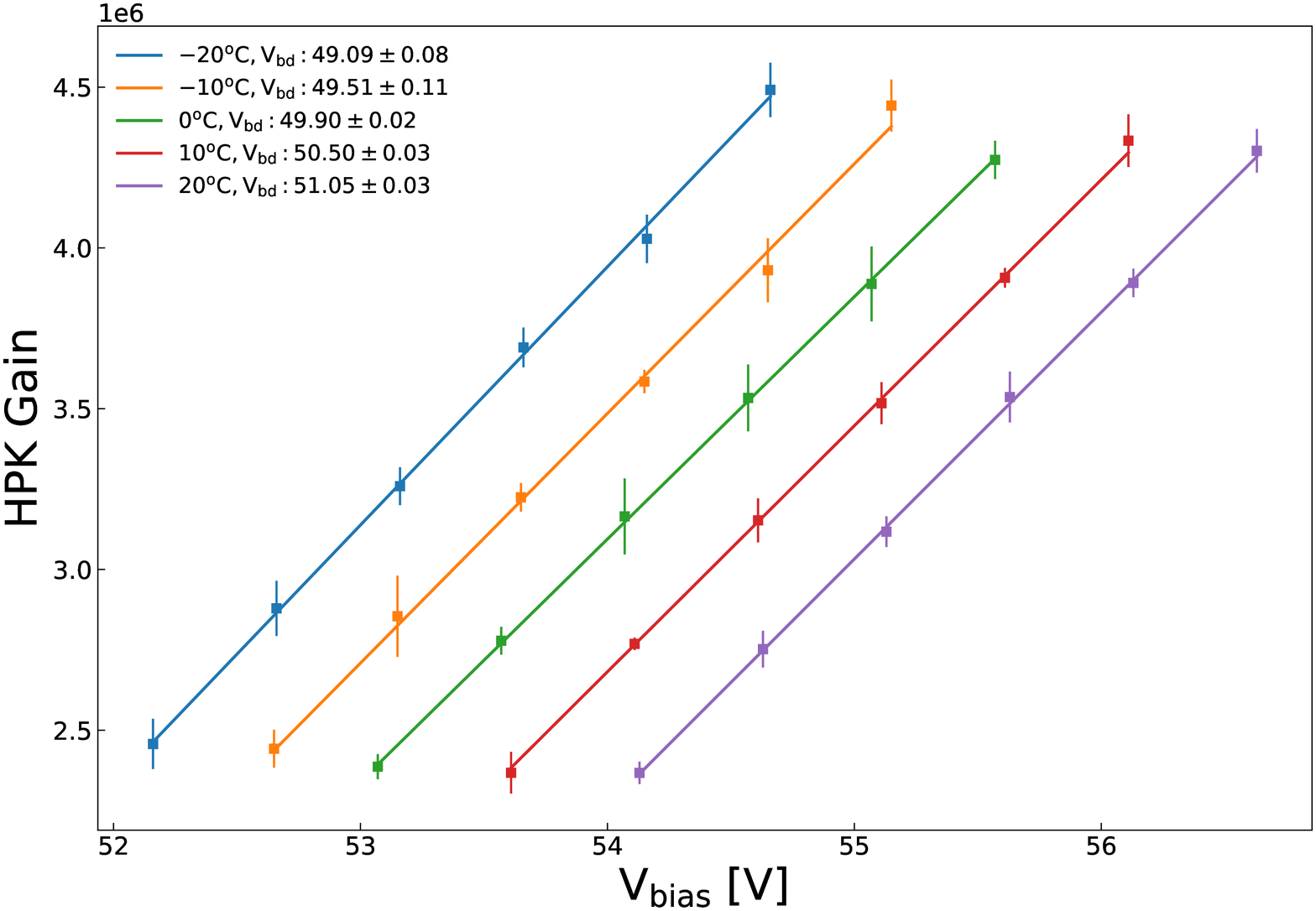}
\hspace{-1cm}
\includegraphics[width=0.55\textwidth]{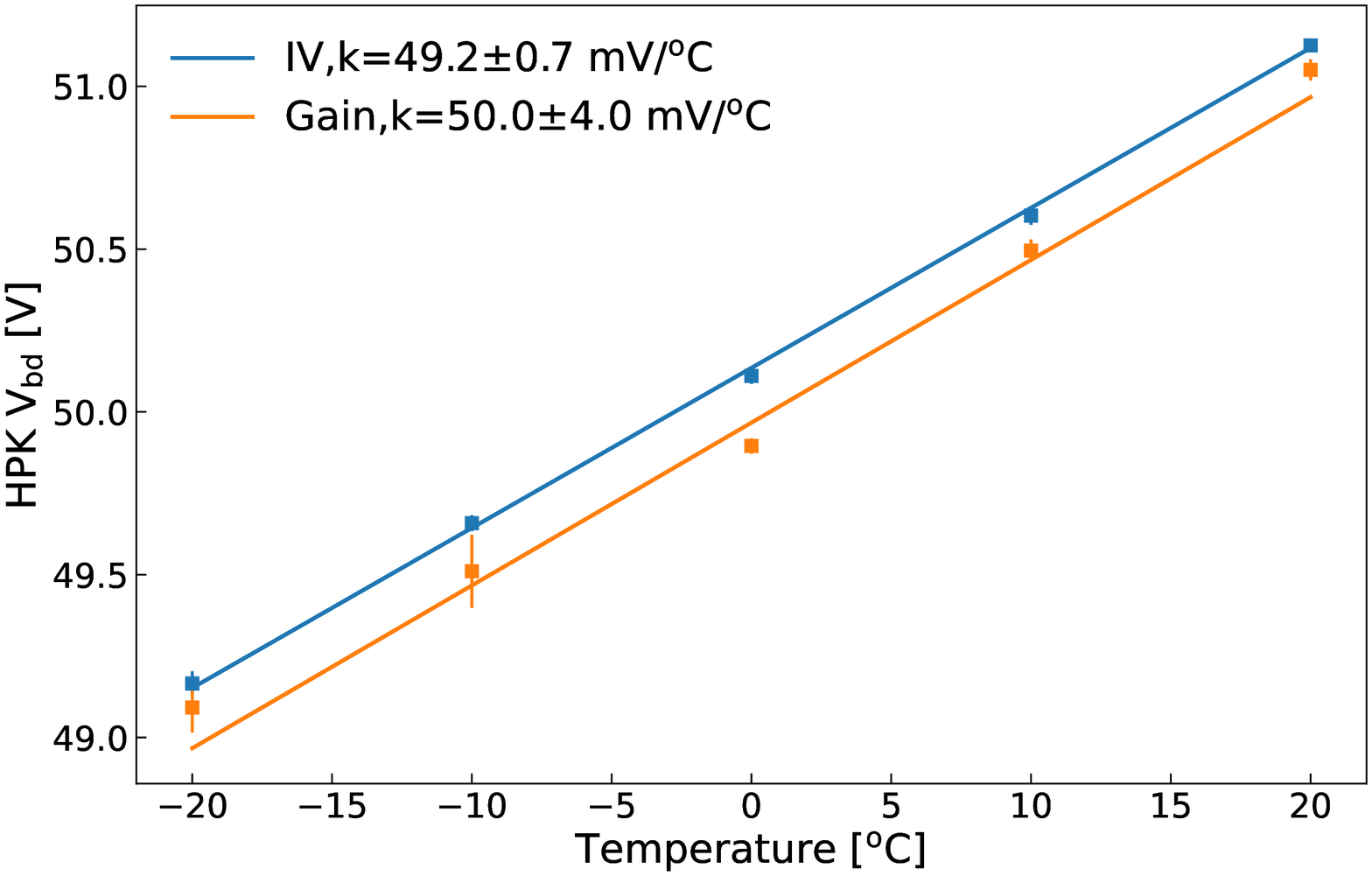} \\
\end{tabular}
\end{center}
\vspace{-0.6cm}
\caption{(Left) Calibrated gain of HPK SiPM operated with different bias voltages at different temperatures. (Right) Comparison of breakdown voltages from gain-voltage method and I-V curves at different temperatures.}
\label{fig:Vbdgain}
\end{figure}

Additionally, the results of breakdown voltages in Fig~\ref{fig:Vbd} are verified by the gain-voltage method ~\cite{Chmill:2017}. The SiPM gain represents the number of carriers in one triggered avalanche and varies with temperature and bias voltage. It can be extracted from the charge spectrum of SiPM waveforms recorded by digitizer DT5742 during the laser on, as shown in Fig.~\ref{fig:QDC}. The difference between two consecutive peaks in the charge spectrum corresponds to the charge $Q$ of a single avalanche.
By fitting peaks with Gaussian distribution in Fig.~\ref{fig:QDC}, $Q$ is calculated as the mean of QDC difference of two consecutive peaks, and the gain is then estimated by:
\begin{equation}
G=\frac{\tau Q}{ A R q_{e}}\;
\end{equation}
where $A$ is the voltage gain of the amplifier, $\tau$ is time step of waveform points in the digitizer, $R$ is the load resistance value of 50 $\Omega$ and $q_{e}$ is the elementary charge. To address the limitation of sampling rate and amplitude resolution of the digitizer, we use the Tektronix MSO5 oscilloscope as a digitizer to calibrate gain from DT5742. Absolute gain results of SiPM are obtained with different bias voltages at different temperatures, as shown in the left Fig.~\ref{fig:Vbdgain}.
The dependence of SiPM gain on bias voltages at a specific temperature is fitted by the formula $G=C\left(V_{\rm bias}-V_{bd}\right)$. The breakdown voltages obtained from this fitting are compared with those derived from I-V curves in the right plot of Fig.~\ref{fig:Vbdgain}, where consistent results of temperature compensation factors are obtained from two methods.

\subsection{Dark noise}
In general, there are two types of dark noise in a SiPM, primary noise and correlated noise. The former is induced by thermally generated carriers or carriers from the band-to-band tunneling effect and is the dominant source of dark noise. The latter is from avalanches subsequent to a primary fired cell from optical crosstalk in neighboring cells due to emitted photons from accelerated carriers or afterpulsing in the same cell due to trapping and subsequent release of carriers. 
\begin{figure}[!t]
\begin{center}
\includegraphics[scale=0.35]{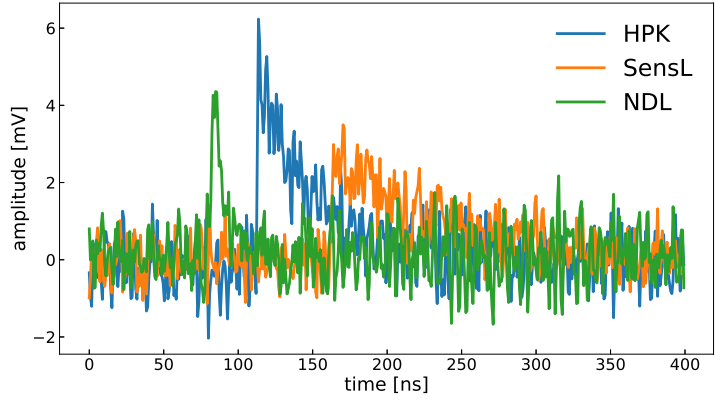}
\vspace{-0.4cm}
\caption{The amplified waveforms of dark noises of 1 p.e. from the HPK, SensL and NDL SiPMs at 20$\rm{^{o}C}$ .}
\label{fig:wave}
\end{center}
\end{figure}

\begin{figure}[!t]
\begin{center}
\includegraphics[scale=0.3]{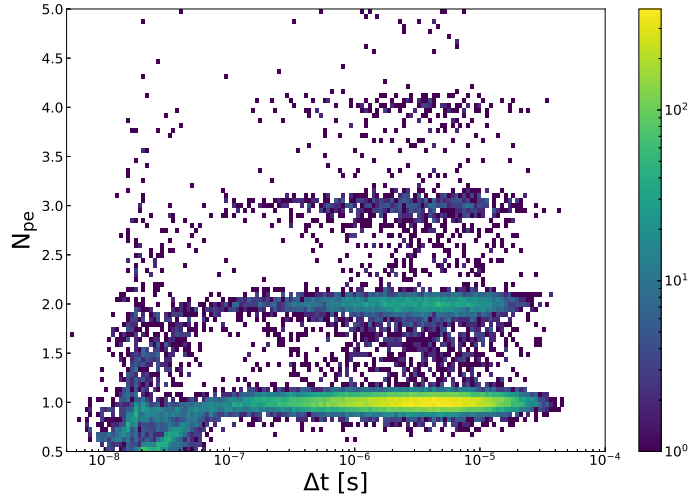}
\vspace{-0.4cm}
\caption{Histogram of peak time interval of two consecutive SiPM signals and the second peak amplitude in photoelectrons. Different colors correspond to different number of events in a bin.}
\label{fig:DCR}
\end{center}
\end{figure}

As illustrated in Fig.~\ref{fig:wave} are the amplified waveforms of dark noises of 1 photoelectron (p.e.) from the three SiPMs at 20$\rm{^{o}C}$. To estimate primary noise rate and crosstalk probability at different temperatures and voltages, we applied a low-pass Butterworth filter and differential leading edge discriminator (DLED) method~\cite{Piemonte:2012rka}. In dark conditions, amplified SiPM waveforms within 1 ms are recorded by the oscilloscope to cover both the primary pulse and its subsequent correlated pulses. These acquired data are firstly smoothed by a low-pass Butterworth filter to reduce fluctuations of the electronic pedestal. Then, they are shifted by 2.4 ns and subtracted from raw waveforms to narrow the pulse width and minimize pile-up signals. Finally, only the rising edge of each signal is preserved while the long falling tail is suppressed. In the updated waveforms, signals are selected with peak amplitude larger than 0.5 photoelectrons. Dark noises from different sources could be identified based on peak amplitudes and time intervals $\Delta t$ between two consecutive signals, as illustrated in Fig.~\ref{fig:DCR}. Note that afterpulsing events with amplitude smaller than one p.e. are partially obscured by pedestal fluctuations of the test setup and not analyzed in this work.

\begin{figure}[!t]
\begin{center}
\begin{tabular}{l}
\hspace{-0.5cm}
\includegraphics[width=0.55\textwidth]{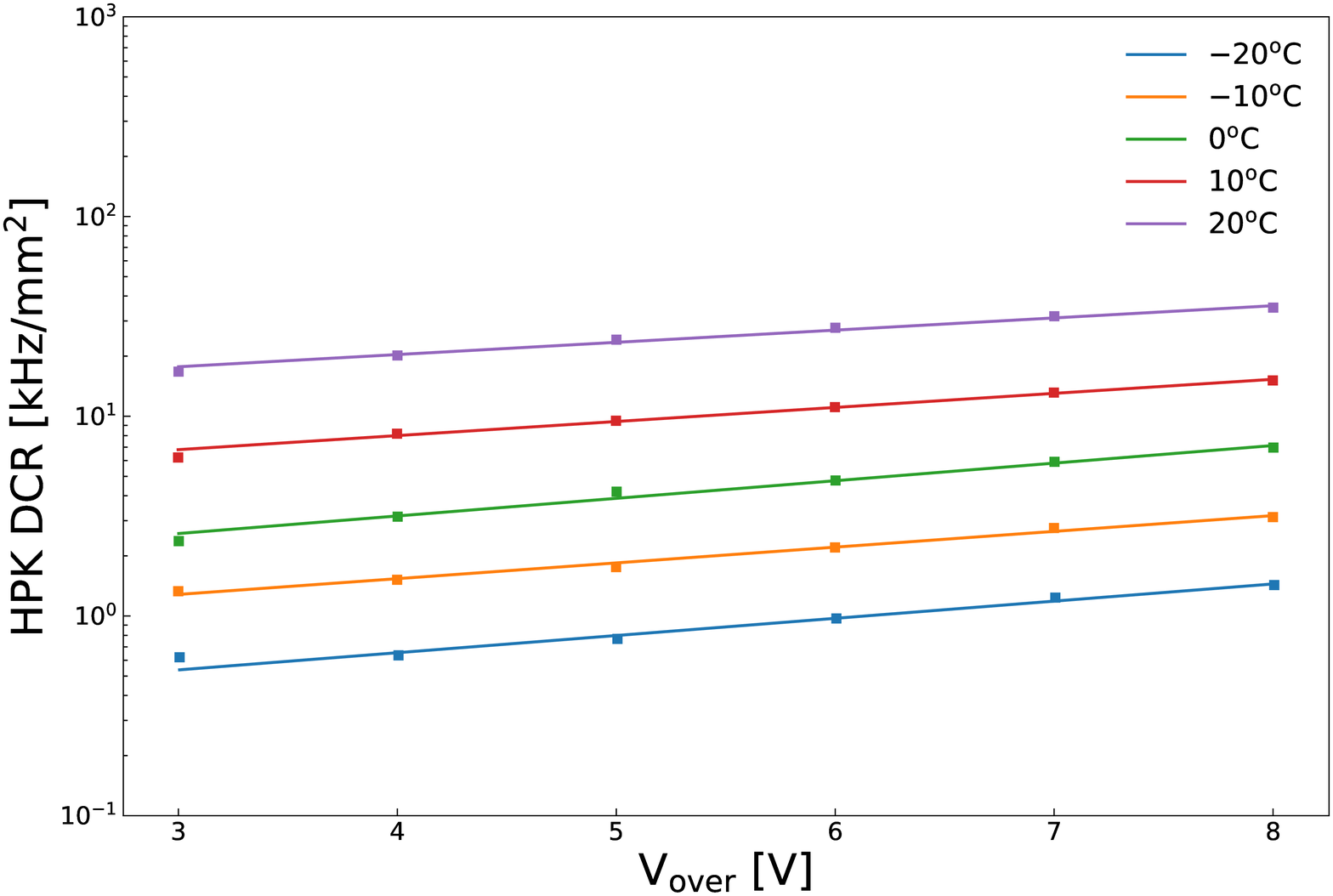}
\hspace{-1cm}
\includegraphics[width=0.55\textwidth]{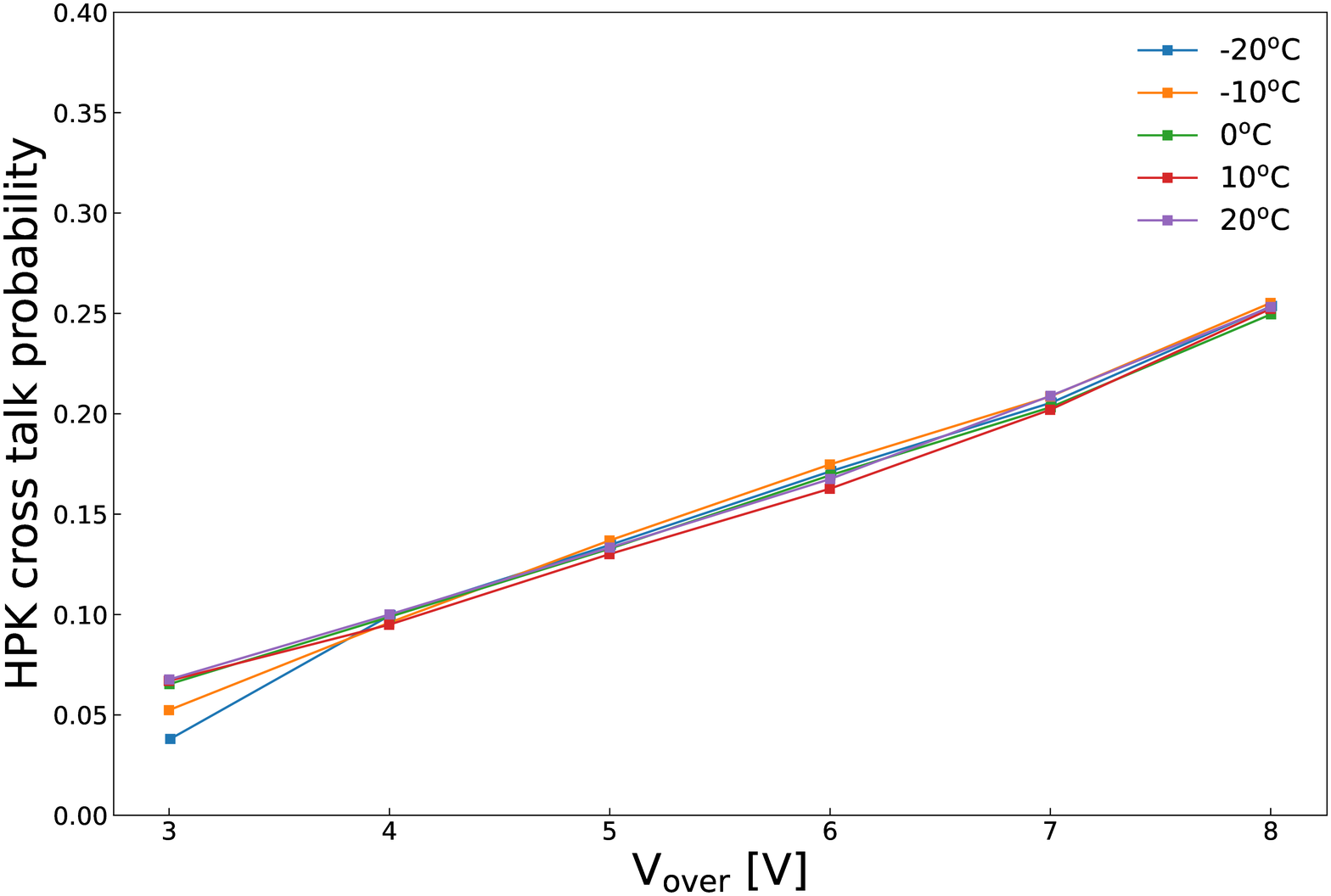} \\
\hspace{-0.5cm}
\includegraphics[width=0.55\textwidth]{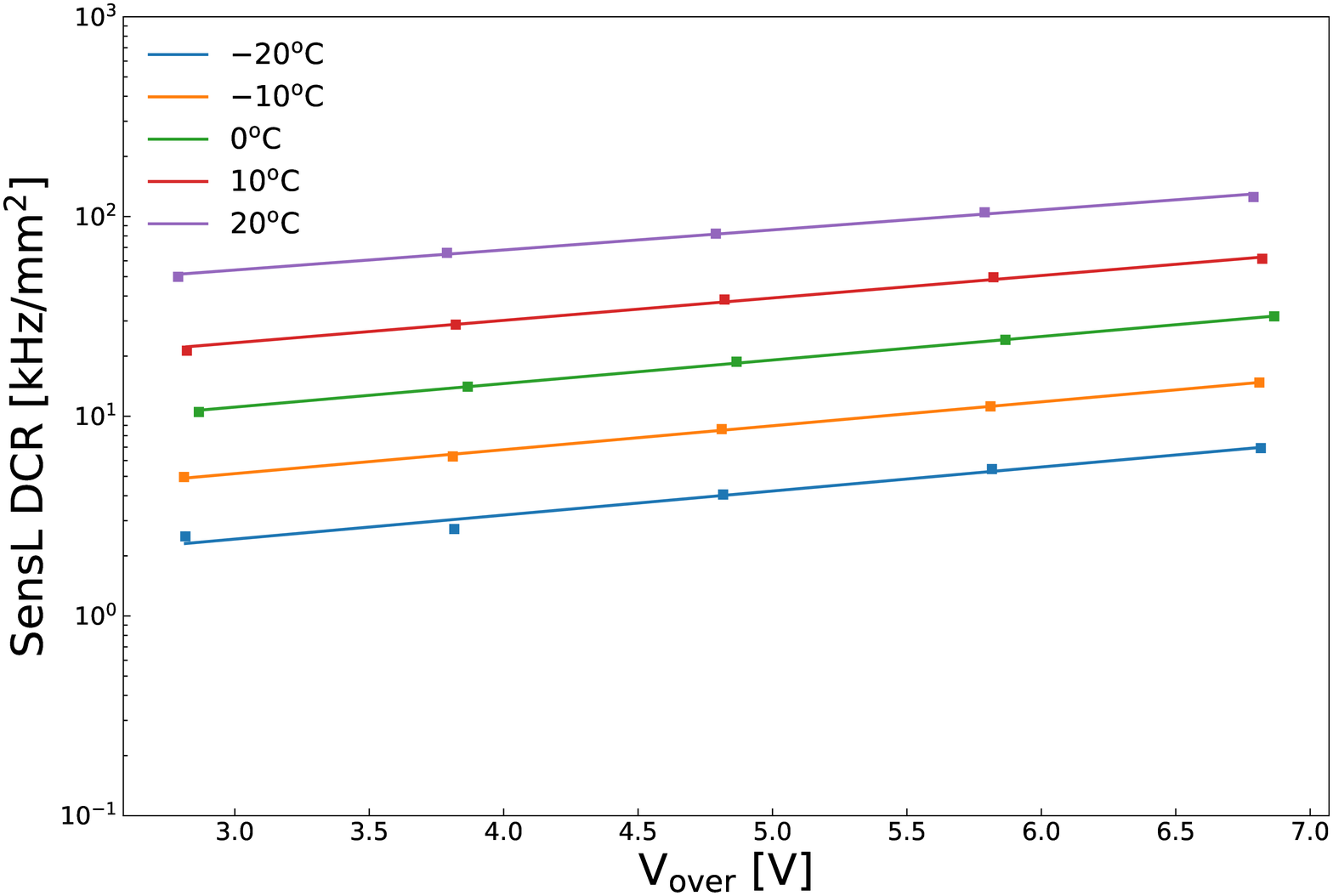}
\hspace{-1cm}
\includegraphics[width=0.55\textwidth]{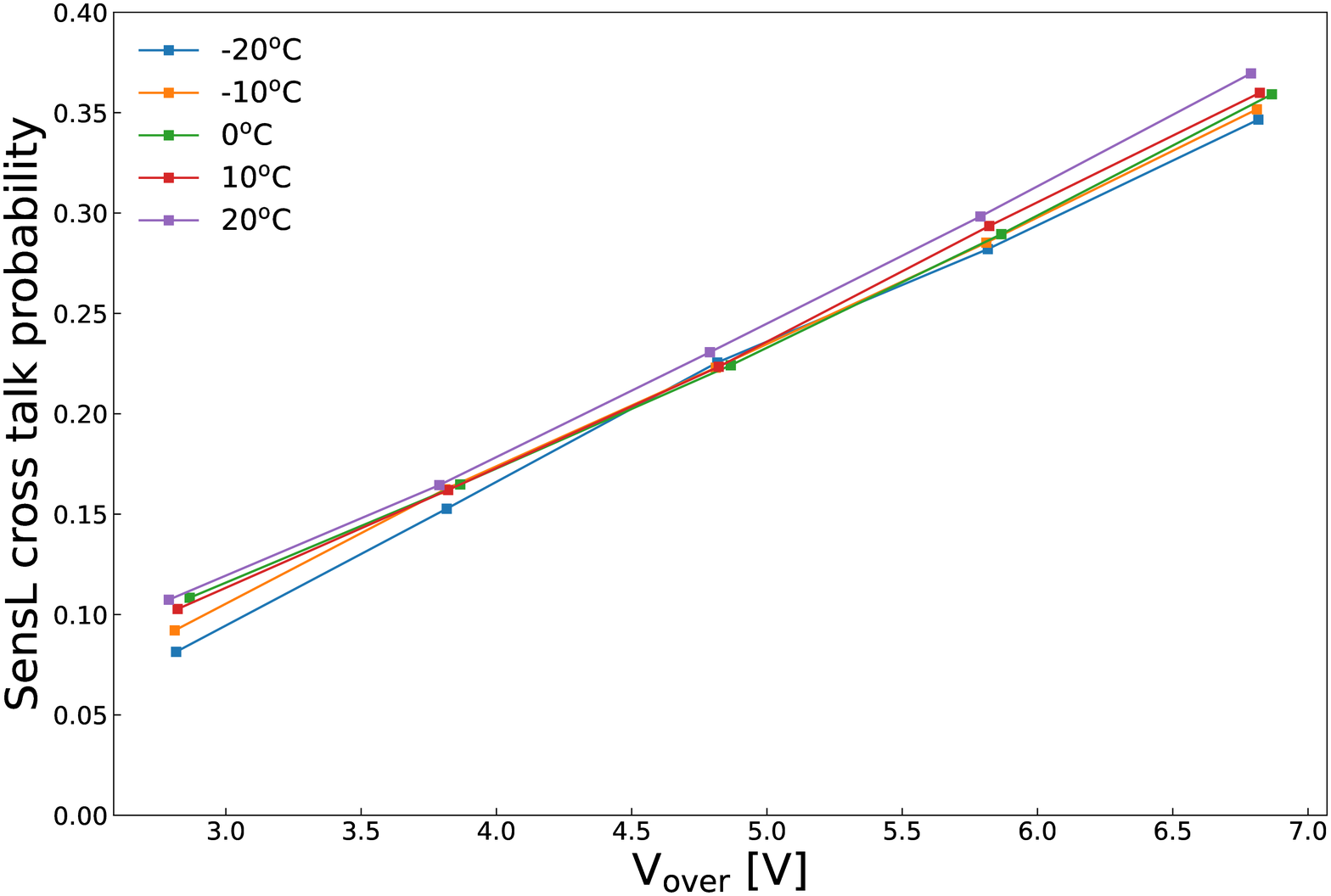} \\
\hspace{-0.5cm}
\includegraphics[width=0.55\textwidth]{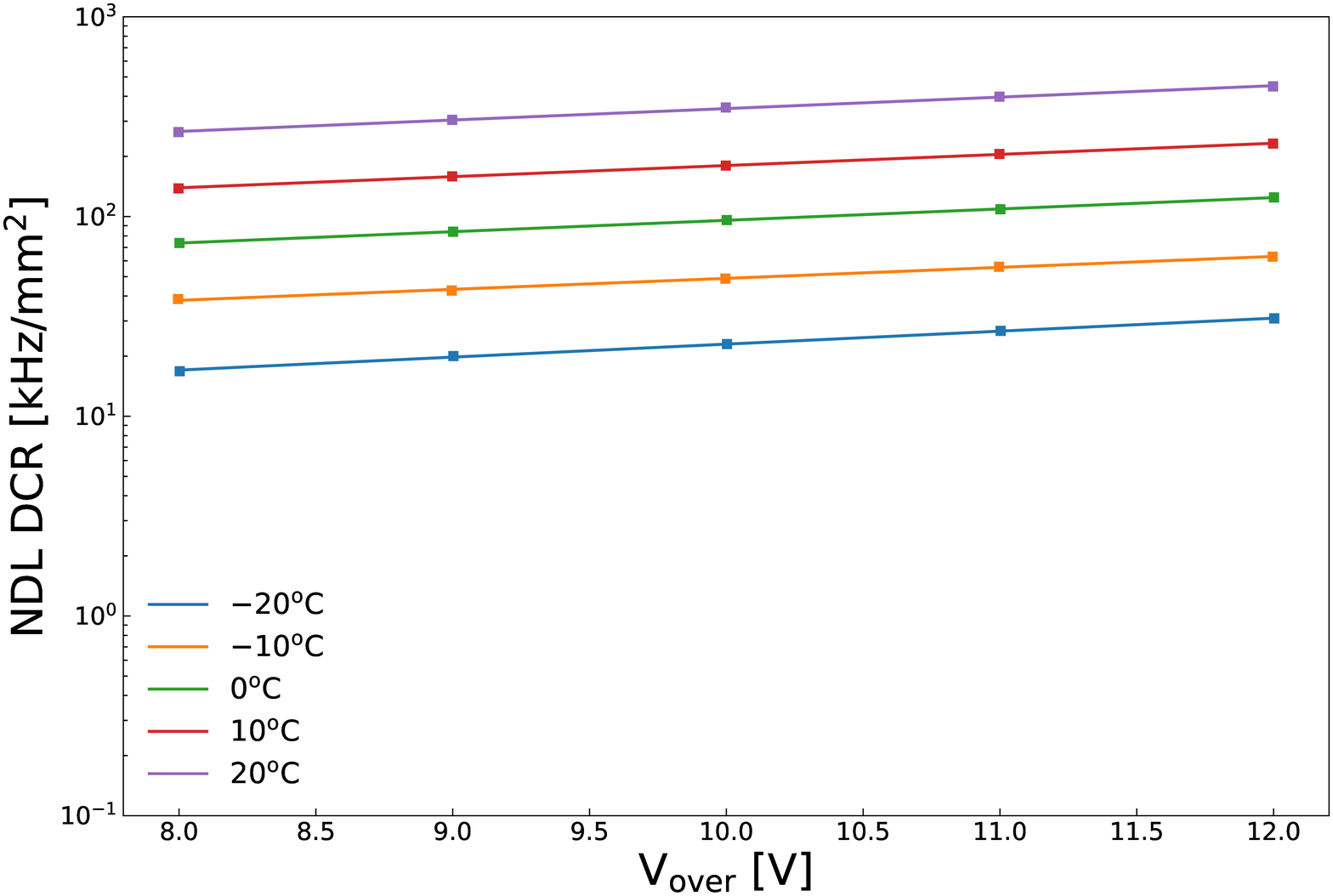}
\hspace{-1cm}
\includegraphics[width=0.55\textwidth]{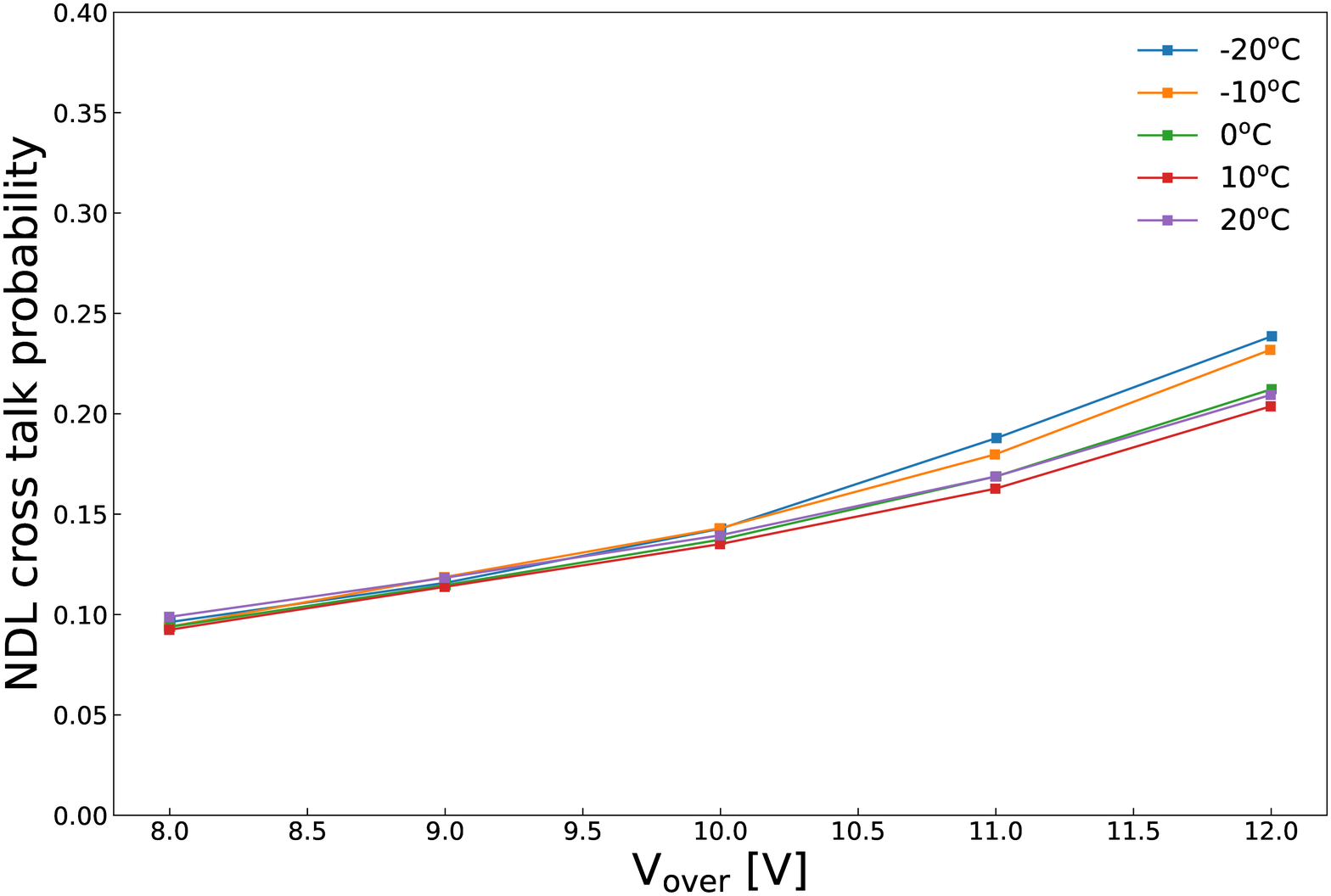} \\
\end{tabular}
\end{center}
\vspace{-0.4cm}
\caption{(Left) Primary dark count rate (DCR) and (Right) crosstalk probability of HPK, SensL and NDL SiPMs operated at different overvoltages and temperatures.}
\label{fig:resultDCRCT}
\end{figure}

Since the primary noises are generated randomly and are independent of each other, the number of these events occurring within a given time interval follows a Poisson distribution. As a result, $\Delta t$ of two consecutive events can be fitted with an exponential formula: 
\begin{equation}
f=r_{d}e^{-r_{d}\Delta t}\;
\end{equation}
where $r_{d}$ refers to primary dark count rate. 
\begin{figure}[!t]
\begin{center}
\begin{tabular}{l}
\hspace{-0.5cm}
\includegraphics[width=0.55\textwidth]{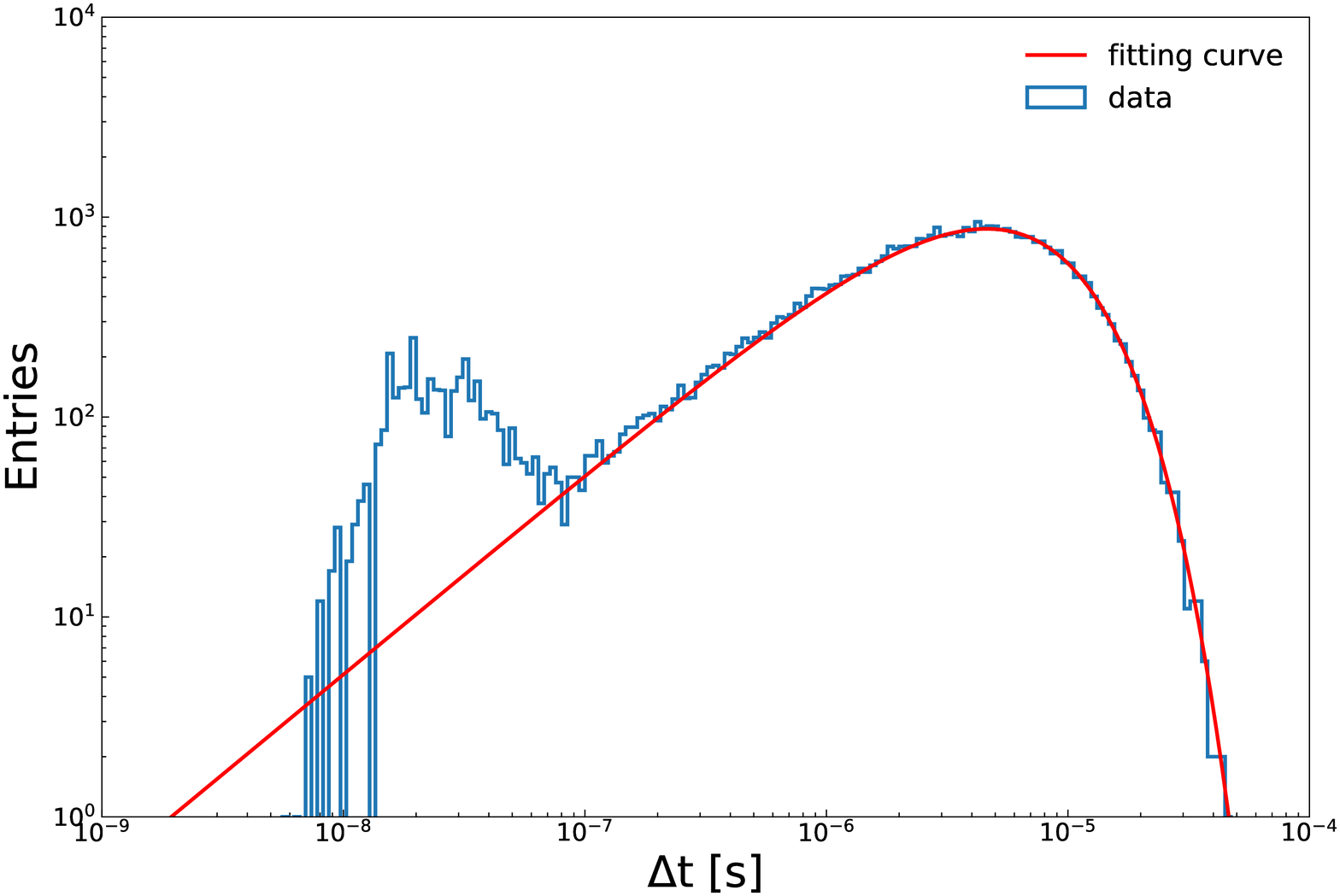}
\hspace{-1cm}
\includegraphics[width=0.55\textwidth]{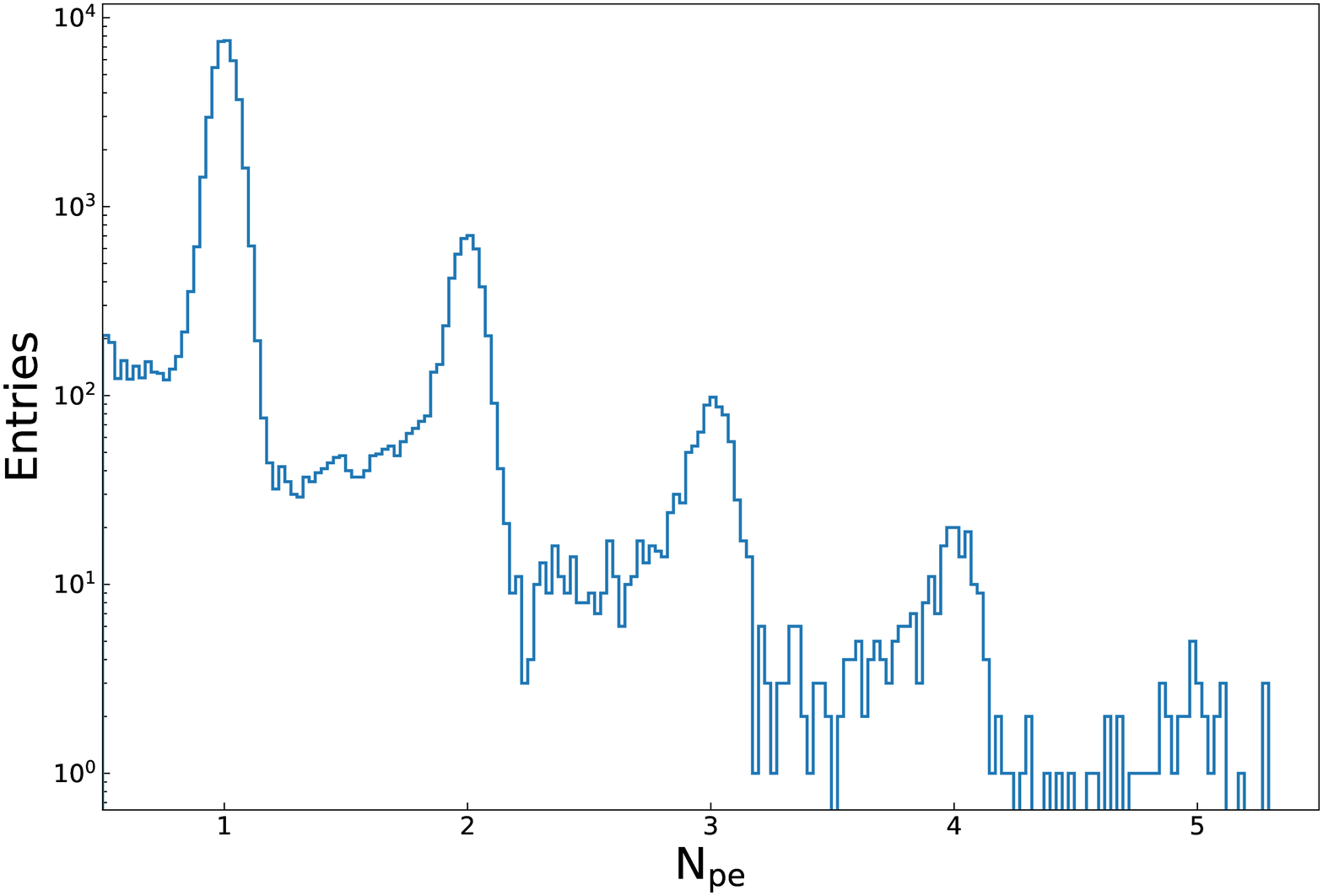} \\
\end{tabular}
\end{center}
\vspace{-0.6cm}
\caption{(Left) Projection of Fig.~\ref{fig:DCR} on $\Delta t$ axis fitted by the exponential distribution. (Right) Projection of Fig.~\ref{fig:DCR} on amplitude axis.}
\label{fig:calDCR}
\end{figure}
The left of Fig.~\ref{fig:calDCR} illustrates the fitting performance of primary noises to the projected of Fig.~\ref{fig:DCR} on time interval $\Delta t$, where the excess events are from delayed crosstalk and afterpulsing. The primary dark count rates per unit area of three SiPMs increase significantly with ambient temperatures and overvoltages ($\rm{V_{over}=V_{bias}-V_{bd}}$) as shown in the left column of Fig.~\ref{fig:resultDCRCT}. The HPK SiPM provides the smallest dark noise rate, while the NDL one has the largest dark noise rate, which partially relates to their different cell pitches.
The optical crosstalk probability subsequent to a primary avalanche is estimated from the right of Fig.~\ref{fig:calDCR}, which is a projection distribution of Fig.~\ref{fig:DCR} on amplitude axis, and defined as the following:
\begin{equation}
CT=\frac{N\left(N_{pe}>1.5 p.e.\right)}{N\left(N_{pe}>0.5 p.e.\right)}\;
\end{equation}
The crosstalk probabilities of three SiPMs are shown on the right of Fig.~\ref{fig:resultDCRCT}, which increase with overvoltages while keeping relatively stable at different temperatures. Note that the relatively lower crosstalk probabilities of -10 $\rm{^{o}C}$ and -20 $\rm{^{o}C}$ at the smallest overvoltage of SensL and HPK are due to events with $N_{pe}$>0.5 p.e. contaminated by pedestal fluctuations.  In general, the crosstalk performance of HPK and NDL SiPM is better than that of SensL at operation overvoltages recommended by manufacturers.

\subsection{Photodetection efficiency}\label{sec:pde}
Photodetection efficiency quantifies the ability of a SiPM to respond to  incoming photons, which depends on the geometrical fill factor of SiPM, quantum efficiency, and the triggering probability of an avalanche in a cell. It is defined as the ratio of the number of detected photons $N_{\rm det}$ to the number of incident photons $N_{\rm inc}$ that reach the SiPM:
\begin{equation}
PDE = \frac{N_{\rm det}}{N_{\rm inc}}\;
\end{equation}

In this work, a photocurrent method with a continuous light source~\cite{Zappala:2016dah} is adopted to evaluate PDE performance at $\rm{20^{o}C}$ of three SiPMs. To suppress the contribution of correlated noises to PDE, we estimate the detected photons by primary noise rate obtained in Fig.~\ref{fig:resultDCRCT}:
\begin{equation}
N_{\rm det} = r_{d}\cdot\frac{I_{\rm sipm}-I^{d}_{\rm sipm}}{I^{d}_{\rm sipm}}
\label{eq:PDE_Ndet}
\end{equation}
where $I_{\rm sipm}$ and $I^{d}_{\rm sipm}$ are SiPM currents in light and dark conditions, respectively. A calibrated PD monitors the incident light of different wavelengths as the following:
\begin{equation}
N_{\rm inc} = 5.03\times10^{15}\cdot\frac{\lambda(I_{\rm pd}-I^{d}_{\rm pd})}{R_{\rm pd}}\cdot\frac{\alpha A_{\rm sipm}}{A_{\rm pd}}
\label{eq:PDE_Ninc}   
\end{equation}
where $I_{\rm pd}$ and $I_{\rm pd}^{d}$ (in unit $A$) are PD currents in light and dark conditions, $R_{\rm pd}$ (in unit $A/W$) is the calibrated responsivity of PD at a wavelength $\lambda$ (in unit $nm$), $A_{\rm sipm}$ and $A_{\rm pd}$ are the active areas of SiPM and the PD, and $\alpha$ is the ratio of light flux at the SiPM position to that at the PD position. 

\begin{figure}[!t]
\begin{center}
\hspace{-0.8cm}
\includegraphics[scale=0.35]{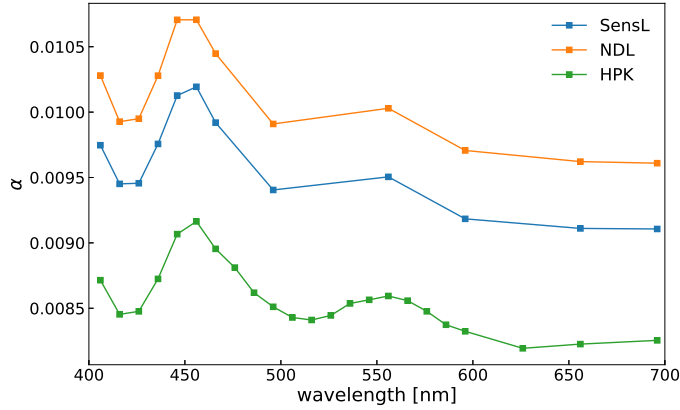}
\vspace{-0.2cm}
\caption{Flux ratios of integrating sphere at different wavelengths for different SiPMs.}
\label{fig:PDEratio}
\end{center}
\end{figure}
Before conducting the PDE measurement, the flux ratios $\alpha$ of the integrating sphere at different wavelengths are evaluated using two calibrated PDs simultaneously. These PDs are positioned at the locations of the SiPM and the PD respectively. Specific adapters are fabricated to ensure that the PD's sensitive plane aligns precisely with the photosensitive surface of the SiPM. A monochromator provides light sources of different wavelengths with a high-power Xenon lamp, of which wavelength values are further verified by a Hamamatsu photonic multichannel analyzer PMA-12. As shown in Fig.~\ref{fig:PDEratio} is the flux ratios at three SiPM positions, where ratio differences at one wavelength of SiPMs are mainly due to position differences of their photosensitive surfaces.   

To achieve an accurate PDE measurement, it is critical to adjust the light intensity reaching the SiPM surface carefully. This adjustment ensures that the SiPM's light current is distinguishable from variations in its dark current while minimizing photon interference with the cells during their recovery time. During PDE measurement, this is realized by adjusting the output slot size of the monochromator to achieve $I_{\rm sipm}$ about 1.5-3 times $I^{d}_{\rm sipm}$ for NDL SiPM, about 3-5 times for SensL SiPM and 10-15 times for HPK SiPM. In this way, the saturation effect of SiPMs is negligible by referring to their primary dark count rates as well as pulse widths.

\begin{figure}[!t]
\begin{center}
\begin{tabular}{l}
\hspace{-0.5cm}
\includegraphics[width=0.55\textwidth]{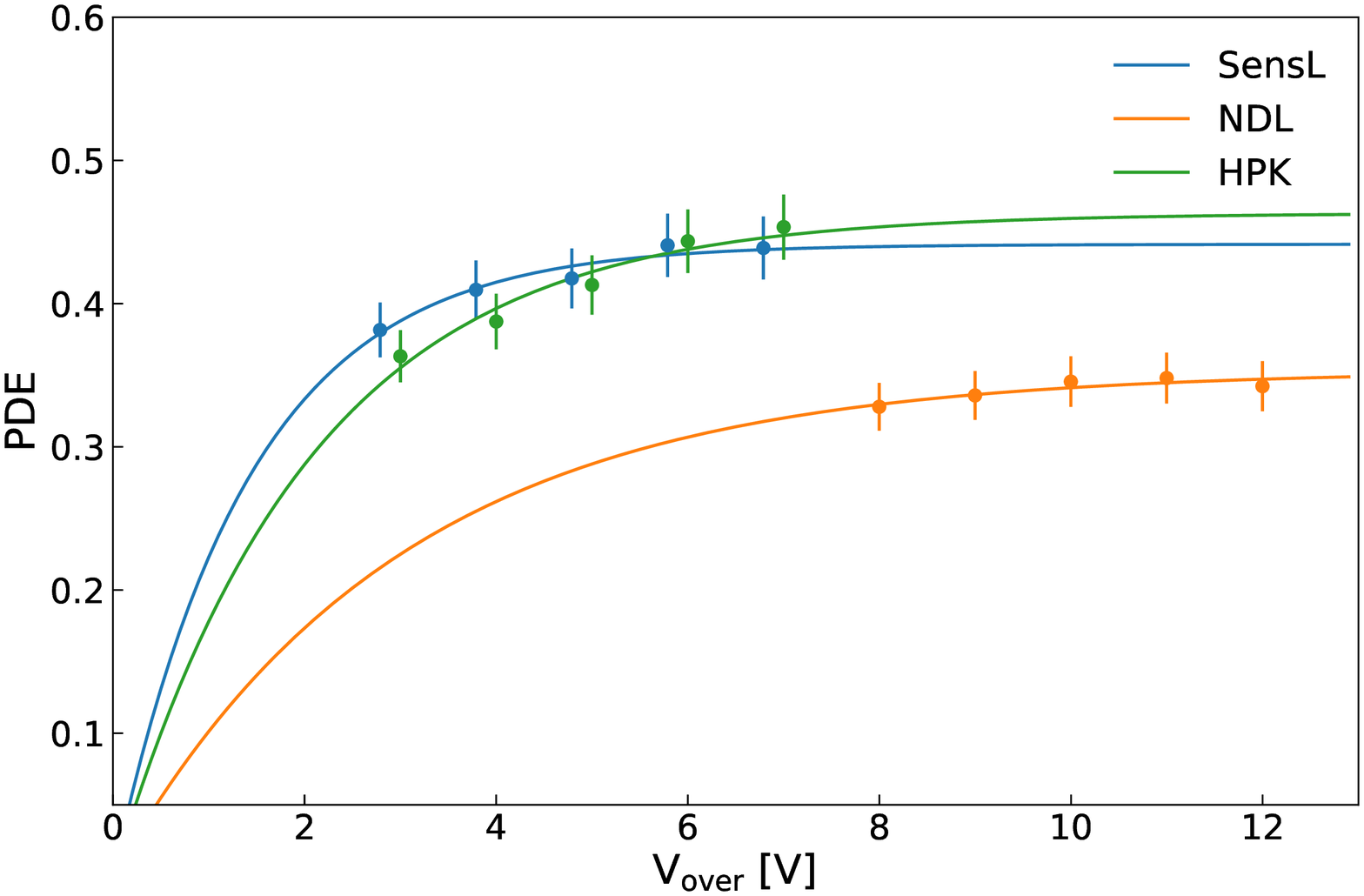}
\hspace{-1cm}
\includegraphics[width=0.55\textwidth]{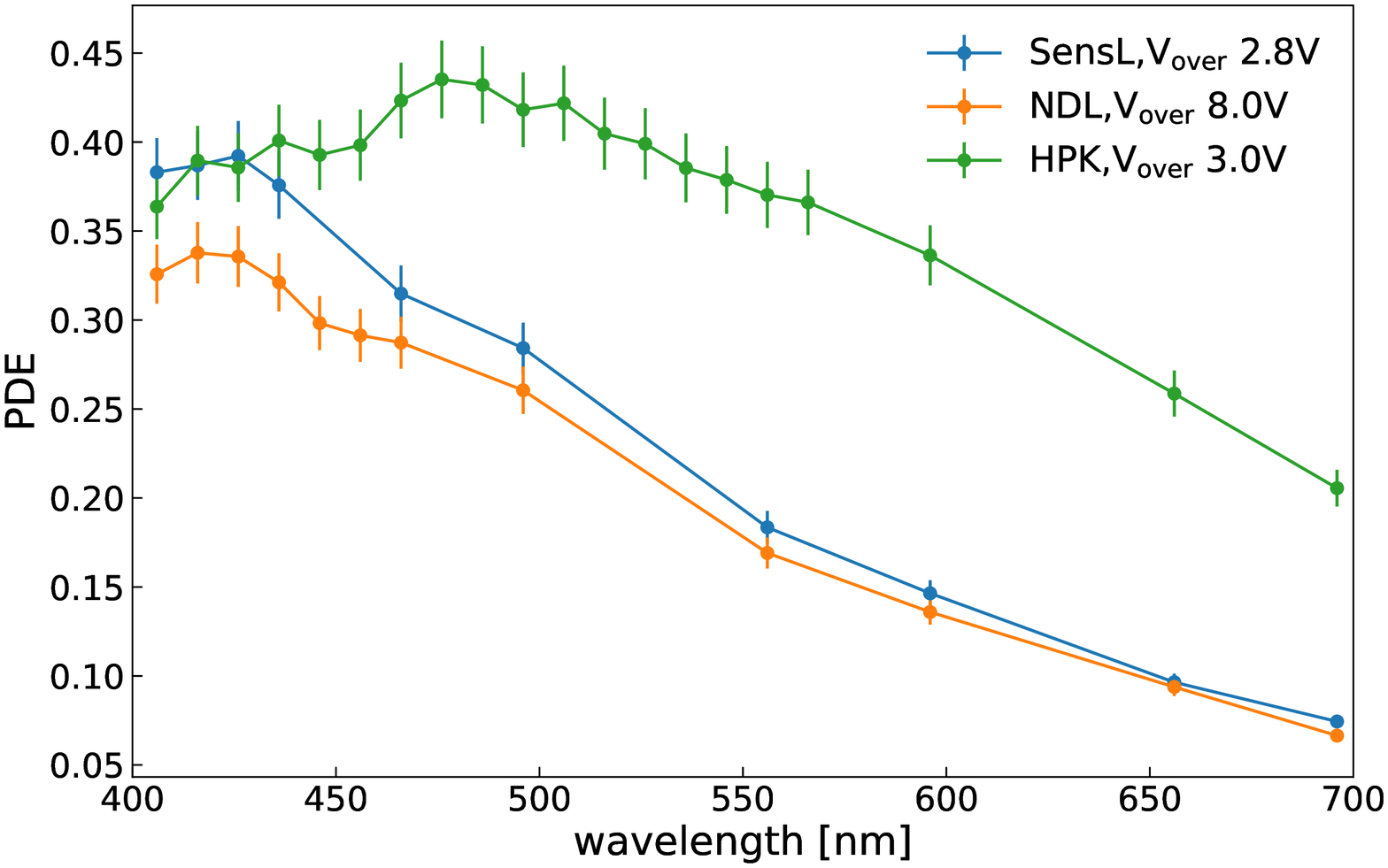} \\
\end{tabular}
\end{center}
\vspace{-0.6cm}
\caption{PDE performance for three SiPMs (left) at 406 nm as a function of overvoltage, and (right) at a fixed overvoltage as a function of wavelength.  }
\label{fig:PDE}
\end{figure}

The PDE measurements are carried out at 20$\rm^{o}C$ with a fixed wavelength as well as a fixed overvoltage for each SiPM as shown in Fig.~\ref{fig:PDE}. The errors of results are dominantly contributed by the 5\% uncertainty of light sources. In the left figure, the PDE of SensL SiPM gets saturated fastest with increase of overvoltages, while that of NDL SiPM is the slowest at 406 nm. In the right figure, the PDE response of HPK SiPM shows more red sensitive than the other two SiPMs, which fits well with emission spectrum from 400 nm to 700 nm of scintillating fibers. 

\section{Conclusion}
In this work, we established a test setup to characterize the breakdown voltages, temperature compensation factors, dark noises, and PDE of three SiPMs from the manufacturers HPK, SensL, and NDL to optimize the performance of the SciFi detector and obtain proper parameters for further SiPM customization. All three SiPMs feature cell pitches that provide a large dynamic range capable of handling photon events from cosmic ray muons in the SciFi detector. However, the HPK SiPM stands out by exhibiting the lowest dark count rate and crosstalk probability as well as the best PDE response at the emission wavelengths of the scintillating fibers. It implies that the best detection efficiency of a SciFi detector with HPK SiPM parameters is expected among these three SiPM products. Therefore, the HPK SiPM will be taken as the reference for further customization of one-dimensional SiPM array for SciFi detector in MST application.

\section{Acknowledgment}
The authors thank Yaoguang Wang for kindly discussions and suggestions. This work was supported in part by the National Natural Science Foundation of China under Grant No.12205174 and by Shandong Provincial Natural Science Foundation under Grant No. ZR2022QA098.

\end{document}